\documentclass[]{jfm}

\usepackage{graphicx}
\usepackage{newtxtext}
\usepackage{newtxmath}
\usepackage{natbib}
\usepackage{hyperref}
\hypersetup{
    colorlinks = true,
    urlcolor   = blue,
    citecolor  = black,
}

\newcommand{\RomanNumeralCaps}[1]
\linenumbers

\usepackage{subcaption}

\usepackage{lscape}

\newcommand{\con}{$CON$}
\newcommand{\KP}{$K2P$}
\newcommand{\intcon}{$\int CON$}
\newcommand{\intKP}{$\int K2P$}
\newcommand{\hotspot}{$\mathbf{x_+}$}

\title{An energetic signature for breaking inception in surface gravity waves}

\author{Daniel G. Boettger \aff{1} \corresp{\email{d.boettger@student.unsw.edu.au}},
  Shane R.  Keating \aff{1},
  Michael L. Banner \aff{1},
  Russel P. Morison \aff{1},
  \and Xavier Barth\'el\'emy \aff{1}}

\affiliation{\aff{1} School of Mathematics and Statistics, University of New South Wales,
Sydney, Australia}

\begin{document}
\maketitle

\begin{abstract}
A dynamical understanding of the physical process of surface gravity wave breaking remains an unresolved problem in fluid dynamics. Conceptually, breaking can be described by \emph{inception} and \emph{onset}, where breaking inception is the initiation of unknown irreversible processes within a wave crest that precede the visible manifestation of breaking onset. In the search for an energetic indicator of breaking inception, we use an ensemble of non-breaking and breaking crests evolving within unsteady wave packets simulated in a numerical wave tank to investigate the evolution of each term in the kinetic energy balance equation. We observe that breaking onset is preceded by around one quarter of a wave period by a rapid increase in the rate of convergence of kinetic energy that triggers an irreversible acceleration of the kinetic energy growth rate. This energetic signature, which is present only for crests that subsequently break, arises when the kinetic energy growth rate exceeds a critical threshold. At this point the additional kinetic energy convergence cannot be offset by converting excess kinetic energy to potential energy or by dissipation through friction. Our results suggest that the ratio of the leading terms of the kinetic energy balance equation at the time of this energetic signature is proportional to the strength of the breaking crest. Hence this energetic inception point both predicts the occurrence of breaking onset and indicates the strength of the breaking event. 
\end{abstract}


\section{Introduction}
The physical process of wave breaking remains one of the classical unresolved problems of fluid dynamics, yet is of fundamental importance for understanding the interaction between the atmosphere and ocean. Wave breaking significantly influences the marine wind drag \citep[e.g.][]{Suzuki2013} and generates enhanced turbulence and energy dissipation in the ocean, modifying the ocean boundary layer over significant depths when coupled with other processes such as Langmuir turbulence \citep{sullivan2007}. The highly nonlinear nature of the breaking process presents challenges for both observational and numerical studies, prompting a range of approaches to develop an objective diagnostic breaking parameter that is valid for any wave type or water depth. \cite{perlin2013breaking-waves-} provides the most recent review of progress in this field and groups diagnostic parameters into three categories that use either the geometric, kinematic or dynamic properties of the wave crest. 

More recently, \cite{Derakhti_2020} introduced the concept of breaking \textit{inception}, which describes the initiation of an irreversible process within the crest that leads to breaking and occurs before the instant of breaking onset, i.e. when the first surface manifestation of breaking occurs at the crest. A diagnostic parameter that is able to characterise breaking inception could therefore provide advance warning of a breaking event and potentially also quantify the breaking strength and the energy dissipated thereafter. A breaking inception parameter may also have broad application to the simulation of wave fields in models where individual breaking events cannot be resolved but in which the energetic processes and dynamic consequences of breaking are important to capture accurately.

The breaking inception indicator proposed by \cite{Derakhti_2020} is based on the diagnostic parameter $B$ \citep{barthelemy2018on-a-unified-br}, which is formally the ratio of the local energy flux to the local energy density normalised by the crest speed $\mathbf{c}$. At the interface, this reduces to the ratio of particle velocity to crest speed $\| \mathbf{u} \| / \| \mathbf{c} \|$. Although this resembles the kinematic breaking criterion in that the value of $B$ at visible breaking initiation is close to unity, it also reveals some remarkable complexities that are yet to be fully understood.  Specifically, \cite{barthelemy2018on-a-unified-br} found that a threshold value $B_{th} = 0.855 \pm 0.05$ exists, beyond which the crest will always evolve to break. This threshold value was subsequently verified in laboratory and computational studies \citep{saket2017on-the-threshol,SAKET2018159}, for a variety of wave packet types \citep{DerakhtiMorteza2018Ptbs}, water depths \citep{seiffert2018simulation-of-b, Derakhti_2020} and in the presence of a constant shear layer \citep{touboul_banner_2021}. 

To avoid ambiguity, \cite{Derakhti_2020} defined  breaking \textit{inception} as the instant at which $B$ first passes through the threshold value $B_{th}$, and characterised breaking \textit{onset} as the instant when visible breaking first occurs. This threshold, which we shall refer to as the kinematic threshold for breaking inception, also provides information on the strength of the subsequent breaking event. \cite{DerakhtiMorteza2018Ptbs} and later \cite{na_chang_lim_2020} found that the normalised rate of change of $B$ as it passes through $B_{th}$, known as $\Gamma$, accurately predicts the breaking strength parameter $b$ \citep{phillips1985}, which has been shown to quantify the energy dissipated through breaking \citep[e.g.][]{drazen2008, Deike2015, sutherland2015}. 

The use of $B_{th}$ is a robust and useful choice as an indicator of breaking inception as it clearly separates breaking and non-breaking crests and can be determined from measurements at the sea surface. However, a dynamical explanation for this threshold value remains elusive and an explanation for why some waves break and others do not requires further investigation. Given that the energetic definition of $B$ reduces to the kinematic diagnostic  $\| \mathbf{u} \| / \| \mathbf{c} \|$, it is possible that $B$ is a proxy variable that accurately distinguishes breaking from non-breaking waves but does not, in itself, track the underlying dynamical cause for breaking inception. 

Dynamic breaking diagnostics have generally focused on the energy growth rate integrated over some region of the crest. \cite{schultz_1994} found that breaking onset could be characterised by the wave integrated potential energy exceeding $52 \%$ of the total energy of the limiting Stokes wave, which suggests that a departure from the equipartitioning of kinetic and potential energy may be a contributor to breaking inception. \cite{song2002on-determining-} constructed an energy growth rate based on the local depth-integrated total energy and the instantaneous wavenumber. This was shown experimentally to distinguish between breaking and non-breaking waves by \cite{banner_2007} but technical challenges with accurately measuring the wavenumber in complex wave packets and a requirement to track the temporal evolution over multiple wave periods have limited its application \citep{barthelemy2018on-a-unified-br}. While these dynamic breaking diagnostics show that the energy growth rate is important to the breaking process, they also demonstrate that accurately and consistently measuring the energetics of an evolving nonlinear wave is nontrivial. Integrating the energy over some subdomain of the wave does ameliorate some of these challenges; however, the energy field is highly focused near the crest tip \citep[e.g.][]{Perlin1996, alberello2018} and the integration process inherently diffuses these local energetic values.

This motivates us to investigate the evolution of the local crest energy field in the time leading up to breaking onset, with the aim of identifying an energetic process that robustly signals breaking inception. Numerical simulation allows us to pursue this in much greater detail than is possible within the constraints of laboratory experiments. We investigate an ensemble of high resolution numerical simulations of non-breaking and breaking wave crests with a range of wave packet sizes and water depths. We track the point with the largest value of local instantaneous kinetic energy, which occurs near the crest tip, and then derive a balance equation for the kinetic energy at this location. We examine the relative contributions of the individual source and sink terms and find that the convergence of kinetic energy at the crest tip provides a reliable indicator of breaking inception a fraction of a wave period prior to breaking onset. The results also indicate a relationship between the convergence of kinetic energy, the rate of change of kinetic energy and the breaking strength parameter $\Gamma$ of \cite{DerakhtiMorteza2018Ptbs}.

\section{Experimental details}

\subsection{Numerical approach}\label{sec:numerical-methods}

We use the Gerris software package \citep{popinet2003gerris:-a-tree-} to generate a suite of numerical simulations of non-breaking, near-breaking and breaking waves across a range of wave packet configurations and grid refinements. Gerris has been extensively validated for simulations of surface gravity waves \citep{WRONISZEWSKI20141}, wave breaking kinematics \citep{deike2017lagrangian-tran,pizzo2016current-generat} and energy dissipation \citep{de2018breaking}. We configure the model to numerically solve the two-dimensional ($\mathbf{x} = (x,z)$), incompressible, variable density Navier-Stokes equations, including the effects of viscosity and surface tension:

\begin{equation}\label{eq:NS}
    \rho \frac{D \mathbf{u}}{Dt} = 
    - \nabla p + \rho \mathbf{g} + \mathbf{f}
    + \mathbf{n} \sigma \kappa \delta_s,
\end{equation}

\begin{equation}\label{eq:mass-balance}
    \frac{D \rho}{D t} = 0,
\end{equation}

\begin{equation}
    \nabla \cdot \mathbf{u} = 0.
\end{equation}

Here, $\rho = \rho(\mathbf{x},t)$ is the fluid density, $\mathbf{u} = (u,w)$ the fluid velocity, $p$ the pressure and $\mathbf{g}$ the gravitational body force. Viscous energy dissipation is characterised by $\mathbf{f} = \nabla \cdot \mu \nabla \mathbf{u}$, where $\mu = \mu(\rho)$ is the dynamic viscosity. The magnitude of the surface tension force is a function of the surface tension coefficient $\sigma$ and the interface curvature $\kappa$, with the force localised at the interface by the Dirac delta $\delta_s$ and the interface normal vector $\mathbf{n}$. Surface tension is modelled through an improved implementation of the continuum-surface-force approach \citep{popinet2009an-accurate-ada} and gravity is applied using the `reduced gravity' method described by  \cite{popinet-2018-sfc-tension} and shown to minimise spurious currents at the interface \citep{WRONISZEWSKI20141}.

A two-phase air-water flow is simulated using the Volume-Of-Fluid (VOF) method, in which the fluid phase is tracked by the conservative tracer $\mathcal{T}$ specifying the fraction of a cell containing water. The limits of  $\mathcal{T} = [0,1]$ indicate that a cell contains purely air and water respectively, while a cell with any intermediate value contains a mixture of the two. We define the location of the air-water interface as the $\mathcal{T} = 0.5$ contour and take the value of scalars on the water-side of the interface as the nearest cell to the interface contour for which $\mathcal{T} = 1$.

\begin{figure}
  \includegraphics[width=\textwidth]{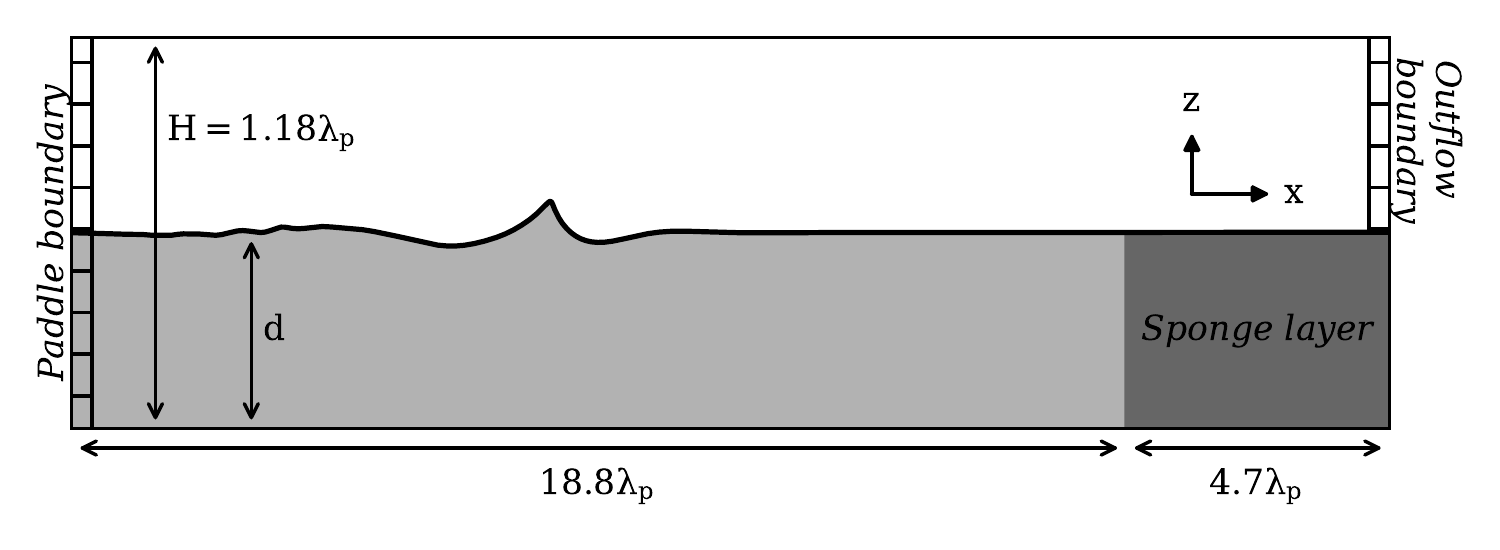}
  \caption{Schematic of our numerical wave tank. Waves are generated at the paddle boundary and travel down the tank in the positive $x$ direction before being absorbed by the numerical sponge layer. A typical chirped wave (enlarged for clarity) is shown, with tank dimensions normalised by the deep-water wavelength $\lambda_p$ derived from the paddle frequency $\omega_p$.}
\label{fig:tank}
\end{figure}

The model is set up as a two-dimensional ($\mathbf{x} = (x,z)$) numerical wave tank in which waves are generated at the left-hand boundary, propagate along the tank and are absorbed at the right-hand boundary (figure \ref{fig:tank}). Previous studies have reported no significant difference in the integrated wave energetics between two- and three-dimensional simulations \citep{de2018breaking, derakhti2016breaking-onset-}, so limiting our study to two-dimensional simulations allows us to examine a wide range of parameters over a large ensemble within computational constraints, while still accurately capturing the energetic characteristics of the waves.

To generate the wave packet, we simulate a bottom-mounted flexible flap paddle by deriving the exact solutions for the velocity and pressure gradient forcing from wavemaker theory \citep{dean1991water-wave-mech} and apply these at the fixed boundary. This method removes the necessity of simulating a moving boundary and thereby greatly increases the computational efficiency of our simulations, while still allowing us to generate a fully nonlinear wave packet. The equivalent lateral movement of the paddle is $<5 \% $ of the wavelength in most cases (table \ref{tab:experiments}) so the approximation of a fixed boundary has little effect on the resultant wave packet. 

The motion of the paddle $x_p$ with time $t$ follows the chirped packet function \citep{song2002on-determining-}
\begin{multline}
x_p(t)=-0.25A_p\left(1+\tanh\left[\frac{4\omega_pt}{N\pi}\right]\right)\left(1-\tanh\left[\frac{4\left(\omega_pt-2N\pi\right)}{N\pi}\right]\right)\\ 
\times\sin\left(\omega_p t\left[1-\frac{\omega_pC_{ch}t}{2}\right]\right)\, 
\end{multline}
where $x_p$ is a function of the paddle forcing amplitude $A_p$, the forcing frequency $\omega_p$, the number of waves in the paddle signal $N$ and the packet linear chirp rate $C_{ch}=1.0112 \times 10^{-2}$. 

The numerical wave tank is configured in non-dimensional coordinates scaled by the linear deep water wavelength $\lambda_p = 2\pi g / \omega_p^2$ and period $T_p = 2\pi / \omega_p$ associated with the paddle forcing frequency $\omega_p$. The height of the tank is $1.18\lambda_p$ with a total length of  $23.5\lambda_p$, the final $4.7\lambda_p$ being configured as a numerical sponge layer. These dimensions allow the wave packet to evolve over at least $18T_p$ after entering the tank, with wave breaking onset generally occurring within half of this time interval. 

Energy absorption at the far end of the tank is achieved through a number of complementary approaches. The final $4.7\lambda_p$ of the tank consists of a numerical sponge layer based on that derived by \cite{clement1996coupling-of-two}, which effectively absorbs high frequency waves. The reflection of low frequency waves is minimised by gradually increasing the grid spacing within the sponge layer to enhance numerical dissipation. An outflow boundary condition is also applied to the dry portion of the lateral boundary to minimise compression of the air phase caused by the paddle motion, which further improves the performance of the model's Poisson solver. 

Gerris uses a quadtree mesh structure that enables efficient adaptive mesh refinement \citep{popinet2003gerris:-a-tree-}. Each level of refinement divides the parent cell into four, resulting in a maximum resolution equivalent to a uniform mesh with $2^n \times 2^n$ grid cells, for $n$ refinement levels. As our primary interest in this study is focused on the air-water interface and the water boundary layer, we determine the maximum required resolution based on the boundary layer thickness $\delta \approx{\lambda_p}/{\sqrt{Re}}$ \citep[eq. (5.7.4)]{batchelor_2000} where $Re=\rho c_p\lambda_p / \mu$ is the wave Reynolds number formulated with the characteristic velocity ($c_p$) and length ($\lambda_p$) scales taken from the paddle signal. To reduce computational cost we set $Re = 4 \times 10^4$ which allows us to resolve the boundary layer with approximately four cells at a refinement level of $2^{10}$ and equates to a resolution of $dx = \lambda_p/870$ with the scaling used. While the wave Reynolds number for a physical deep water gravity wave is $Re \approx 1 \times 10^6$, previous studies \citep{deike2017lagrangian-tran, mostert_deike_2020} have shown that $Re = 4 \times 10^4$ is large enough that viscous effects are not dominant and all energy within the boundary layer is adequately resolved. We also conducted a limited number of experiments with a maximum refinement level of $2^{11}$ (approximately eight cells within the boundary layer, equivalent to $dx = \lambda_p/1740$) to confirm that the total energy of the simulation did not change (see appendix \ref{sec:Verification}). For all experiments, mesh refinement criteria are configured to ensure maximum resolution at the air-water interface and in regions of large vorticity.

\subsection{The crest ensemble}\label{sec:ensemble}

We use our numerical wave tank to conduct a suite of simulations across a range of wave packet configurations, water depths and grid resolutions (table \ref{tab:experiments}). Each individual crest in the wave packet is tracked in space and time (as described below) with the evolution of the crest geometry and energetics recorded. To account for the finite resolution of our numerical simulations, we characterise a crest as breaking if the interface contour exceeds the vertical by a horizontal distance $d\eta_x \ge 0.5 {dx}$ over a length $d\eta_z \ge {dx}$ where $dx$ is the finest model grid scale. We use the qualitative term ``near-breaking'' to describe the steepest and most energetic non-breaking crests for which the local interface contour closely approaches vertical but does not exceed it. The local crest energetics are measured at the location $\mathbf{x_+} = [x_+(t), z_+(t)]$ where the local kinetic energy density $E_k$ has its maximum value. A crest reference location and time are set as $[x_0, t_0] = [x_+, t]$ at the instant of breaking onset for breaking crests and at the instant of maximum local $E_k$ for non-breaking crests. The evolution of the crest in space and time is then referenced to these parameters using the non-dimensional coordinates $x^* = (x - x_0) / \lambda_p$,  $z^* = z / \lambda_p$ and $t^* = (t - t_0) / T_p$.

Cubic interpolation is used to determine the location of \hotspot{} and the value of the energetic quantities at this point, but the unsteady movement of the crest \citep[e.g.][]{Derakhti_2020, fedele_2020} does lead to a level of uncertainty in the resultant time-series. This is managed by filtering the data with a running mean of width $0.15T_p$, which we find effectively removes high-frequency noise without significantly smoothing the temporal variability in the data. Peak values at breaking onset are preserved by applying the filter independently for $t^* \le 0$ and $t^* > 0$ (i.e. before and after breaking onset) with the window width gradually reducing to zero for $\vert t^* \vert < 0.15T_p$. The difference between the original and the filtered data is used to estimate the $5 \%$, $95 \%$ confidence interval using a bootstrap method. We use these confidence intervals to objectively discard crests for which the energetic parameters presented in this study are not correctly captured by our analysis methods, which usually occurs when a crest is impacted by droplets from another breaking crest in the same wave packet. Any crests for which the relative magnitude of the confidence intervals exceed the $95$th percentile for that parameter are discarded, which accounted for $7.5\%$ of the total. The final ensemble consists of $581$ non-breaking and $73$ breaking crests (table \ref{tab:experiments}). 

\begin{figure}
    \begin{subfigure}[b]{0.5\textwidth}
         \centering
         \includegraphics[width=\textwidth]{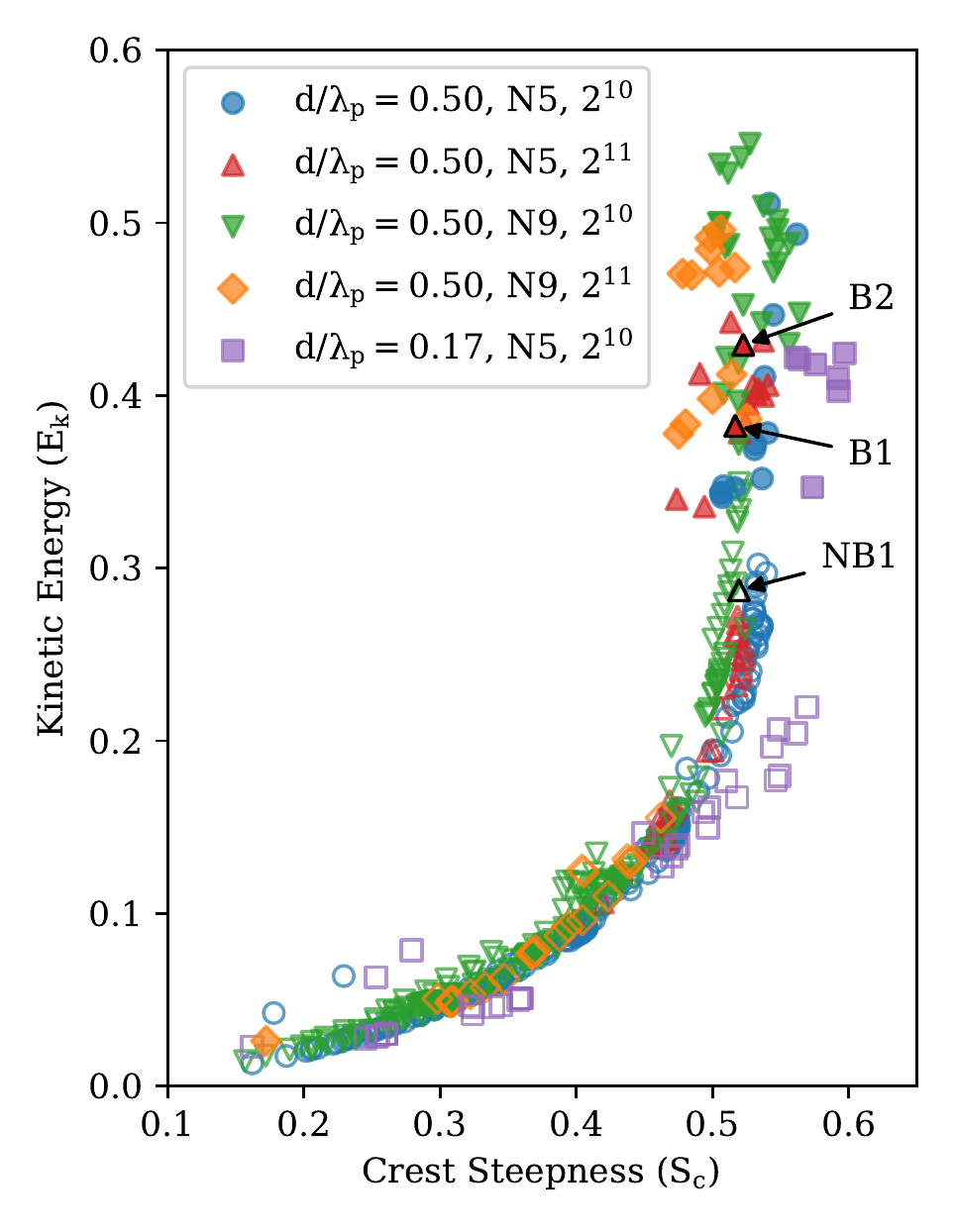}
    \end{subfigure}
    \begin{subfigure}[b]{0.5\textwidth}
         \centering
         \includegraphics[width=\textwidth]{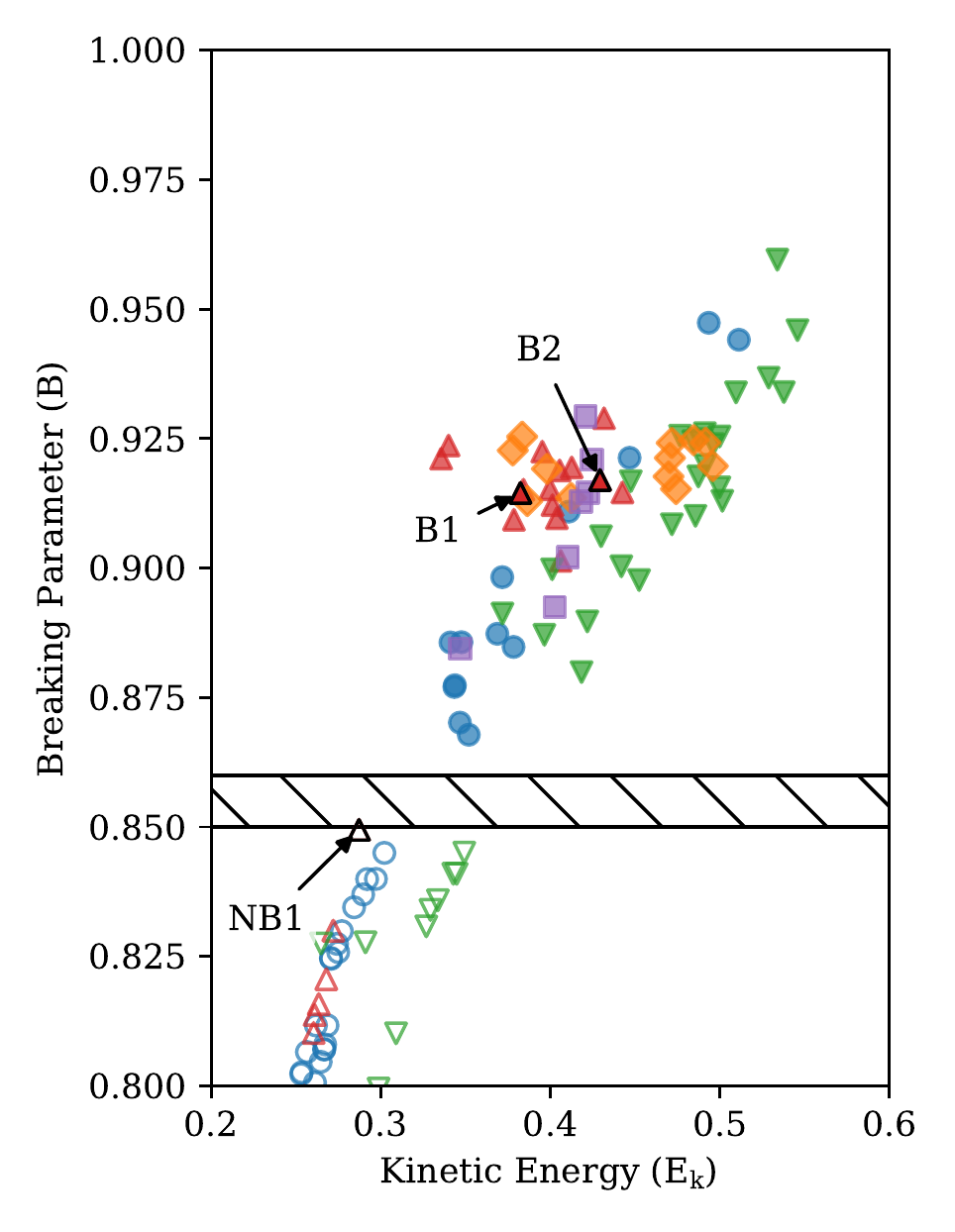}
    \end{subfigure}
  \caption{Summary of the individual crests included in the ensemble. The model was configured using a range of mesh refinement levels ($2^x$), wave packets $N$ and water depth $d/\lambda_p$. For each configuration the amplitude of the paddle $A_p/\lambda_p$ was varied to generate an ensemble of crests with different energy levels. Non-breaking (breaking) crests are indicated by hollow (filled) symbols. The representative crests shown in subsequent figures are labelled NB1, B1 and B2. (Left) The maximum kinetic energy $E_k$ as a function of local wave steepness $S_c$. (Right) The value of $B$ at $t^*=0$ (non-breaking crests with $B < 0.8$ not shown); the threshold $B_{th} = 0.855 \pm 0.05$ reported by \cite{barthelemy2018on-a-unified-br} is shown by the hatched region.}
\label{fig:KE-and-B}
\end{figure}

In figure \ref{fig:KE-and-B} (left panel), we characterise this ensemble in terms of the local $E_k$ at \hotspot{} and the crest steepness $S_c = \pi a / \lambda_c$, which captures the unsteady and time-dependant development of the crest in terms of the amplitude $a$ and zero-crossing wavelength $\lambda_c$ \citep{banner2014linking-reduced}. Figure \ref{fig:KE-and-B} (left) illustrates the distinct energetic characteristics of each wave packet type: for a given $E_k$, deep-water $N=5$ crests are typically steeper than $N=9$ crests, and $N=5$ crests in intermediate depth are even steeper. However, the local $E_k$ is not sufficient to distinguish breaking from non-breaking crests, with a mix of both cases occurring in the range $0.3 < E_k < 0.35$.  

We also examine the ensemble in terms of the breaking inception parameter $B$. We find that the location \hotspot{} of maximum particle velocity and kinetic energy is found on the forward face of the crest, corresponding with previous observational \citep{Perlin1996} and viscous numerical \citep{VARING2020103755} studies. This is in contrast to the inviscid simulations of \cite{barthelemy2018on-a-unified-br}, who chose to follow the crest tip $\mathbf{x_c}$. We find that the location \hotspot{} may be offset by up to $0.015\lambda_c$ from $\mathbf{x_c}$ and with a particle velocity $\Vert \mathbf{u}_+ \Vert$ up to $10 \%$ greater than the crest tip particle velocity $\mathbf{u_c}$ at the time of breaking onset, with these differences generally being larger for more energetic crests. We also note that in a recent $B_{th}$ validation study by \cite{Derakhti_2020}, the particle velocity was taken as the maximum value within $\approx 0.03 \lambda_c$ of the crest tip. This motivates us to construct $B$ as 

\begin{equation}\label{eq:B-alternate}
    B = \frac{\Vert \mathbf{u}_+ \Vert}{ \Vert \mathbf{b}_+ \Vert},
\end{equation}

\noindent where the denominator $\mathbf{b_+}=d \mathbf{x_+} / d t$ is chosen primarily to be consistent with the mathematical formulation outlined in the following sections. We find that the threshold $B_{th} = 0.855 \pm 0.05$ is replicated by our ensemble (figure \ref{fig:KE-and-B}, right) using (\ref{eq:B-alternate}). We also found (not shown) that the same threshold value is also observed if the crest speed $\mathbf{c}$ is used in the denominator of (\ref{eq:B-alternate}), but that $B$ is underestimated if $\Vert \mathbf{u_c} \Vert$ is used as the numerator. Because the observed difference between $\Vert \mathbf{b}_+\Vert$ and $\Vert \mathbf{c} \Vert$ is less than $\pm 5 \%$ we see that $B$ is relatively insensitive to the choice of crest velocity, although we do note that $\mathbf{b}$ varies more smoothly in time than $\mathbf{c}$ as the latter can change rapidly if the crest is impacted by surface ripples that make identification of $\mathbf{x_c}$ challenging. With this measure, figure \ref{fig:KE-and-B} (right) shows that our ensemble covers the threshold region between breaking and non-breaking crests with both varying wave packets and water depths. 

\begin{table}[h]
  \begin{center}
\def~{\hphantom{0}}
  \begin{tabular}{ c c c c c c c }
  Refinement & & & & Total & \multicolumn{2}{c}{Total crests} \\
    level    & $N$  & $d/\lambda_p$ & $A_p/\lambda_p$   & simulations & Breaking & Non-breaking \\[3pt]
    $2^{10}$     	    & 5     & 0.59          & $0.025-0.05$   & 44    & 13    & 249		    \\
    $2^{10}$            & 9     & 0.59          & $0.025-0.045$   & 49    & 26    & 234  	        \\
    $2^{10}$            & 5     & 0.20          & $0.08 - 0.092$ & 9     & 6     & 34            \\
    $2^{11}$            & 5     & 0.59          & $0.037-0.043$   & 19    & 15    & 40 		     \\
    $2^{11}$            & 9     & 0.59          & $0.037-0.0389$   & 9	    & 13    & 24	         \\
  \end{tabular}
  \caption{Summary of experiments included in this study. The model was configured using a range of mesh refinement levels, wave packet size $N$ and water depth $d/\lambda_p$. For each configuration the amplitude of the paddle $A_p/\lambda_p$ was varied to generate an ensemble of breaking and non-breaking crest cases.}
  \label{tab:experiments} 
  \end{center}
\end{table}

\begin{figure}
 \includegraphics[width=\textwidth]{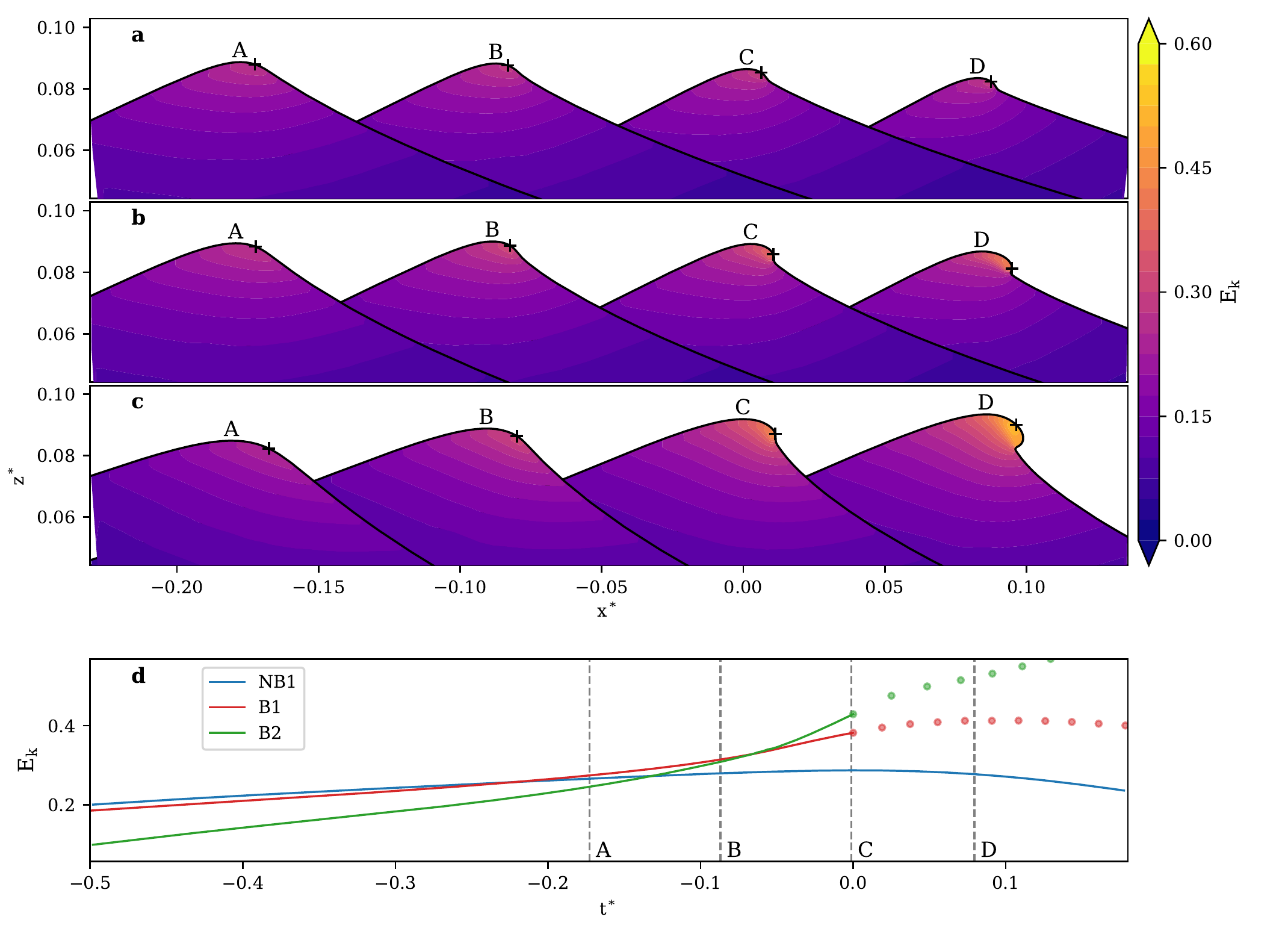}
  \caption{The evolution of $E_k$ for the (a) NB1 (near-breaking), (b) B1 and (c) B2 (breaking) representative crests as they progress through the growing and decaying phase. Snapshots are equally spaced at intervals of $0.09T_0$ and the vertical axis is exaggerated by a factor of $7:1$. The $+$ indicates the location $\mathbf{x_+}$ where $E_k$ has its maximum value, and in (d) the temporal evolution of $E_k$ at this location is shown. Periods of active breaking are indicated in panel (d) by the dotted lines. The time of each snapshot (A-D) corresponds with the vertical dashed lines. Snapshot $C$ occurs at the time that $E_k$ at $\mathbf{x_+}$ has its maximum value for the near-breaking crest, and the time of breaking onset for breaking crests.}
\label{fig:KE-slice}
\end{figure}

We have selected representative near-breaking (NB1) and breaking (B1, B2) crests from the deep-water $N=5$ cases that span the $E_k$ and $B$ parameter space (figure \ref{fig:KE-and-B}) and use these in subsequent sections to illustrate the key features of the crest energetics. Some initial observations of the characteristics of these crests can be made from figure \ref{fig:KE-slice}, in which the evolution of the local $E_k$ field for each crest is shown. The NB1 crest is the near-breaking case in our ensemble that most closely approaches breaking. As the crest grows, a distinct bulge develops on the crest tip and the local interface angle becomes near-vertical. This region is also associated with a local concentration of elevated $E_k$. The B1 case is a weakly breaking crest in which the interface only briefly exceeds the vertical before again relaxing. Conversely, the B2 case illustrates a stronger breaking example in which more extensive overturning of the interface is evident. 

If only the shape of these crests were considered it could be concluded that the separation between near-breaking and breaking is simply a function of a marginal increase in steepness which eventually leads to the local interface angle exceeding the vertical. However, the values of $E_k$ at the location of the maxima \hotspot{} (figure \ref{fig:KE-slice}d) demonstrate that the energetics of these crests follow diverging paths. Until $t^* \approx -0.15$ the values of $E_k$ are similar in all three cases, but at this point the $E_k$ in the near-breaking case reaches a plateau and begins to gently decrease, while the $E_k$ in the breaking cases undergoes rapid increase up to breaking onset ($t^*=0$) and beyond. The spatial extent of this increase in $E_k$ is seen in the snapshots of the crest evolution (figure \ref{fig:KE-slice}b-c), with the region of intensifying $E_k$ magnitude fully encompassing the formation of the crest tip bulge. It is evident from this that the convergence of $E_k$ within the crest tip is an important factor in the breaking process. This motivates our analysis presented in the follow sections, in which we mathematically describe and quantitatively track the evolution of this process. 

\section{Evolution of the crest energetics}\label{sec:energetics}

\subsection{Mathematical formulation}

To examine the crest energetics we first construct a balance equation for the local kinetic energy density $E_k = \tfrac{1}{2} \rho \vert\mathbf{u}\vert^2$. This is derived by taking the scalar product of (\ref{eq:NS}) with the fluid velocity $\mathbf{u}$ and making use of $\nabla \cdot \mathbf{u} = 0$ to obtain
\begin{equation}\label{eq:Ek-balance}
    \frac{D E_k}{Dt} = 
    \frac{\partial E_k}{\partial t} + \mathbf{u} \cdot \nabla E_k = 
    -\nabla \cdot \mathbf{u} p 
    - \rho g w
    + \mathbf{u} \cdot \mathbf{f}
    + \mathbf{u} \cdot \mathbf{n} \sigma \kappa \delta_s, 
\end{equation}
where $w$ is the vertical particle velocity. The terms on the right hand side account for the work against pressure, the work against gravity, viscous energy dissipation and the surface tension force respectively. 

A similar balance equation for the potential energy density $E_p = \rho g \zeta$ is derived by multiplying the mass balance equation \eqref{eq:mass-balance} by $\mathbf{g}$ and the particle vertical displacement $\zeta$ to give
\begin{equation}\label{eq:Ep-balance}
    \frac{D E_p}{Dt} = 
    \frac{\partial E_p}{\partial t} +\mathbf{u} \cdot \nabla E_p = 
    \rho g w.
\end{equation}
The $\rho g w$ term links (\ref{eq:Ek-balance}) and (\ref{eq:Ep-balance}) and quantifies the exchange  between $E_k$ and $E_p$: as water particles are advected upwards (downwards) $E_k$ is lost (gained) and $E_p$ is gained (lost) at equal rates. Because the $\rho g w$ term depends only on the vertical component of the particle velocity, the efficiency of this energy conversion process relative to the other terms in (\ref{eq:Ek-balance}) depends on the direction of the velocity vector $\mathbf{u}$, which varies in space and time as the wave evolves. 

The Lagrangian balance equations (\ref{eq:Ek-balance}) and (\ref{eq:Ep-balance}) provide useful insights into the energetics following a fluid parcel; however, their interpretation as a function of time is not useful in this context as the fluid particles follow an orbital motion that does not correspond with the geometric evolution of the crest. A more insightful approach results from defining a location that moves and evolves with the crest geometry. This may be a specific location such as the highest point of the crest $\mathbf{x_c}$, or the location where a scalar quantity has its maximum value. In our case, we choose to follow the location \hotspot{} of maximum $E_k$ as it is situated on the forward face of the crest tip, where breaking onset is first observed (figure \ref{fig:KE-slice}). This location also varies smoothly in time, which aids the interpretation of the kinetic energy evolution. When following the location \hotspot{}, the change in $E_k$ has both a local and a convective component described by the operator $D_b E_k /Dt = \partial E_k/ \partial t + \mathbf{b_+} \cdot \nabla E_k$ \citep[][eq. 2.2]{tulin-2007}, where $\mathbf{b_+}=d \mathbf{x_+} / d t$. When applied to (\ref{eq:Ek-balance}) this leads to
\begin{equation}\label{eq:DbEkDt-balance}
\frac{D_b E_k}{D t}= \underbrace{- \nabla \cdot \left(\mathbf{u} p + [\mathbf{u-b_+}]E_k \right)}_{\text{\con{}}} 
    \underbrace{- \rho g w}_{\text{\KP{}}} 
    + \underbrace{\mathbf{u \cdot f}}_{\text{friction}} + \underbrace{\mathbf{u \cdot n} \sigma \kappa \delta_s}_{\text{sfc tension}}
\end{equation}
where the relevant kinetic energy flux velocity is seen to be $\mathbf{u - b_+}$. This result can also be derived by considering the rate of change of kinetic energy within an arbitrarily small control volume (see Appendix \ref{sec:sensitivity}, where an application of this approach is shown explicitly). We subsequently refer to the terms on the right hand side as the convergence term ($CON$), the kinetic to potential energy conversion term ($K2P$), friction and surface tension. The use of (\ref{eq:DbEkDt-balance}) allows us to track these energetic quantities and relevant source terms at our location of interest on the evolving crest. While the magnitude of kinetic energy does not discriminate between breaking and non-breaking crests (figure \ref{fig:KE-and-B}), we show in the subsequent section that by tracking the location of maximum $E_k$ we also track the location where the leading terms in (\ref{eq:DbEkDt-balance}) have their maximum values. 

\subsection{Temporal evolution}

\begin{figure}
 \includegraphics[width=\textwidth]{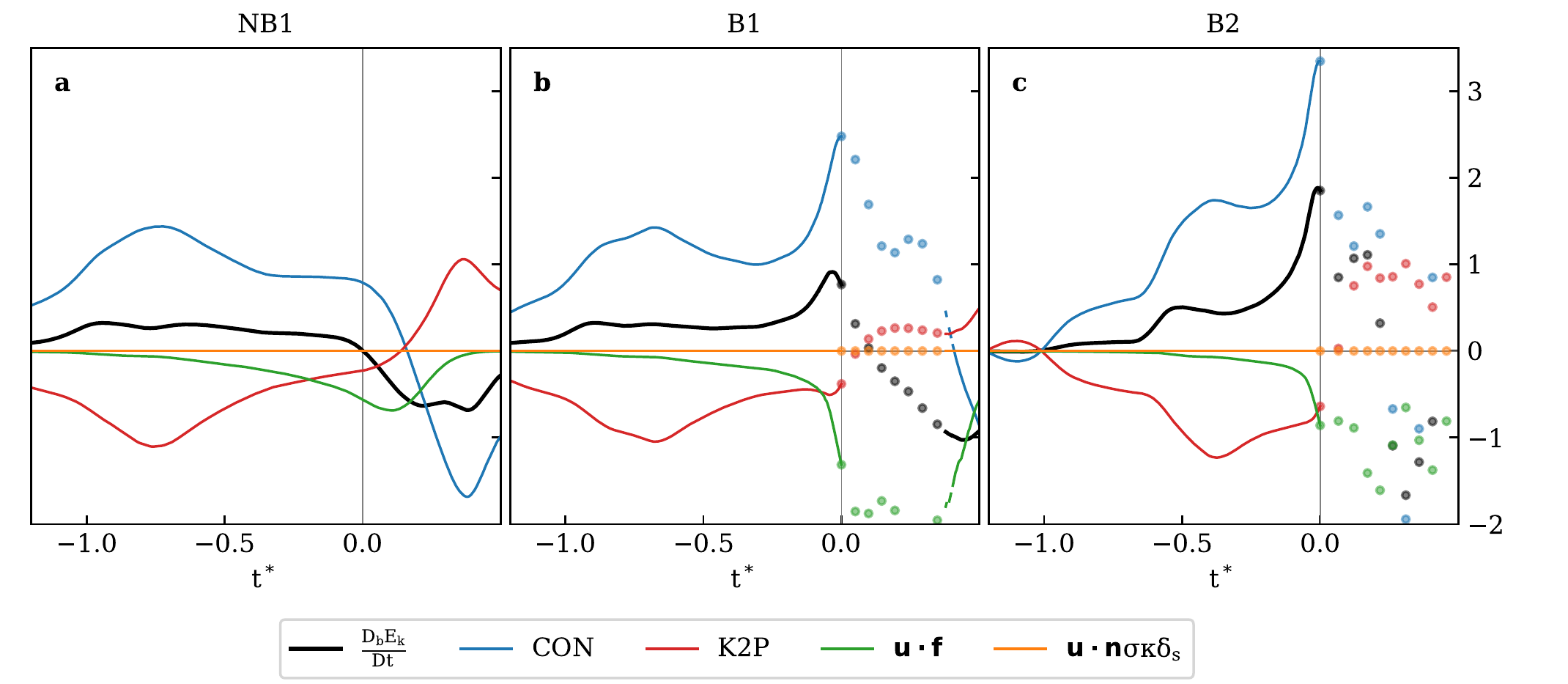}
    \caption{The rate of change of the local kinetic energy density $E_k$ at the location \hotspot{} (figure \ref{fig:KE-slice}) and its contributing terms from (\ref{eq:DbEkDt-balance}) for the representative near-breaking (NB1) and breaking (B1, B2) crests. Positive (negative) values indicate a source (sink) of $E_k$. Periods of wave breaking are indicated by the solid dots.}
\label{fig:local-energetics}
\end{figure}

We examine $D_b E_k / D t$ and its components in figure \ref{fig:local-energetics}, where each term of (\ref{eq:DbEkDt-balance}) is represented as a kinetic energy source / sink by applying the appropriate sign to the values, so that positive (negative) values represent an increase (decrease) in kinetic energy. 

For all cases, the kinetic energy generally increases ($D_b E_k / D t > 0$) to reach a peak value around $t^* \approx 0$ (i.e. when $E_k$ reaches its maximum value for the non-breaking cases or when breaking onset occurs for breaking cases) before decreasing ($D_b E_k / D t < 0$) as $t^* > 0$. 
The evolution of $D_b E_k / D t$ is dominated by the convergence term \con{} and the kinetic to potential energy conversion term \KP{}, with surface tension of negligible magnitude and friction significant only near $t^*=0$. The \con{} and \KP{} terms are of nearly equal magnitude and opposite sign, such that when one is acting as a source of $E_k$ the other is a sink. 

The near-cancellation of these two terms is observed throughout the evolution of the near-breaking case NB1 and is also seen throughout most of the growth phase of the breaking cases B1, B2. But a striking deviation from this balance develops as the B1 and B2 crests approach breaking onset. From $t^* \approx -0.1$, a rapid increase in \con{} is seen which is not balanced by a corresponding increase in \KP{}. As a result, $E_k$ also increases rapidly up to breaking onset. Unlike the \con{} term, the \KP{} term for all three example crests are of similar magnitude and trajectory. This indicates that for a given crest there is a limit to the amount of $E_k$ that can be converted to $E_p$. The excess convergence of kinetic energy results in the development of the visible $E_k$ hotspot in figure \ref{fig:KE-slice} and the subsequent breaking event. 


\begin{figure}
	\includegraphics[width=\textwidth]{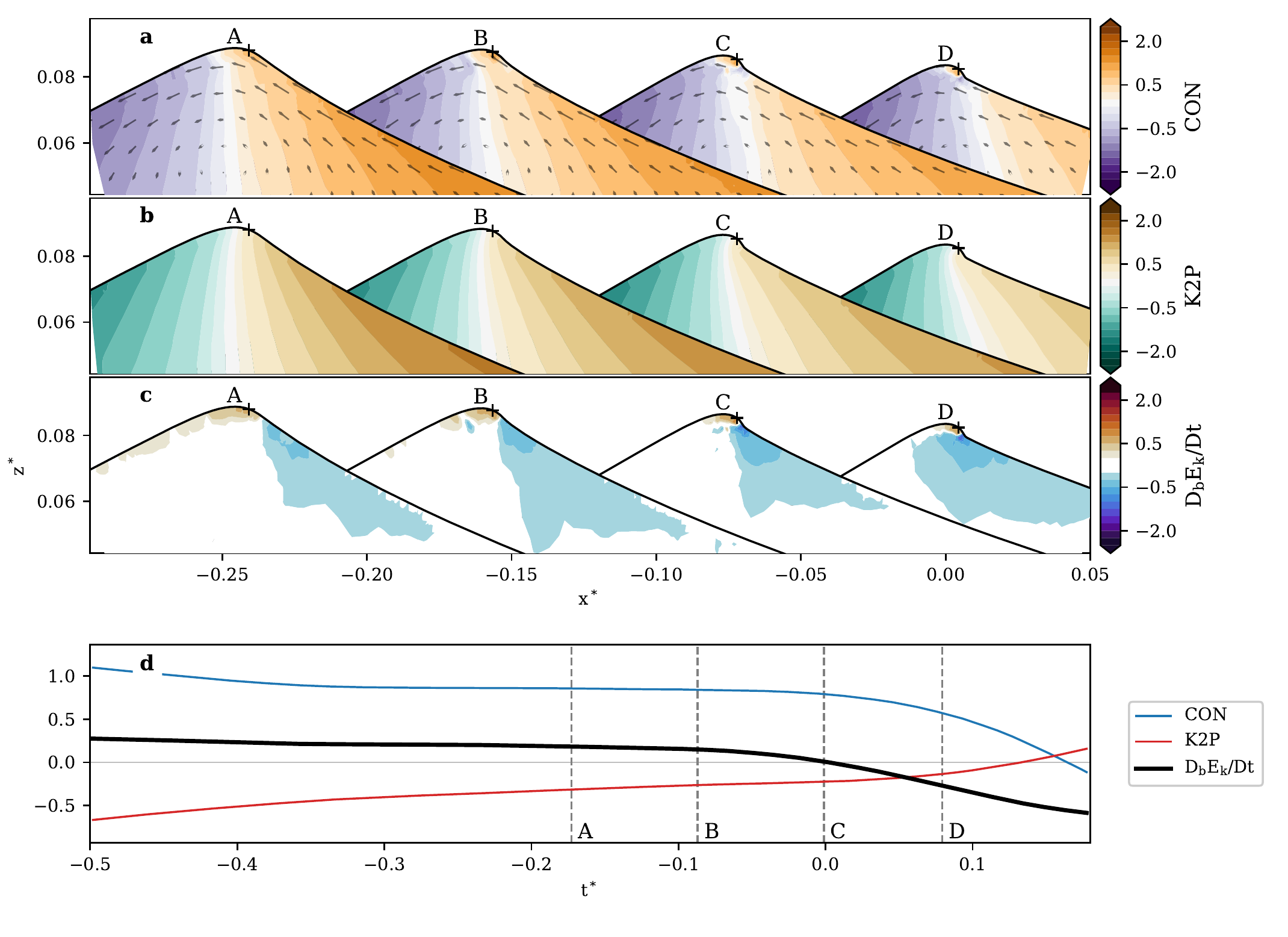}
  \caption{The evolution of the kinetic energy balance for the NB1 crest. The dominant terms of the local balance equation (\ref{eq:DbEkDt-balance}) are (a) \con{}, (b) \KP{} and (c) $D_b E_k / Dt$. Vectors in (a) show the relative magnitude of the $\mathbf{u - b}$ flux velocity at each snapshot. The $+$ indicates the location $\mathbf{x_+}$ where $E_k$ has its maximum instantaneous value. Snapshots are equally spaced at intervals of $0.09T_0$ and the vertical axis is exaggerated by a factor of $7:1$, with snapshot $C$ occurring at the time that $E_k$ at $\mathbf{x_+}$ reaches its maximum value. The temporal evolution of each term in (a), (b) and (c) at $\mathbf{x_+}$ have been taken from figure \ref{fig:local-energetics} and shown in (d) for comparison.}
\label{fig:balance-slice-NB1}
\end{figure}

\begin{figure}
	\includegraphics[width=\textwidth]{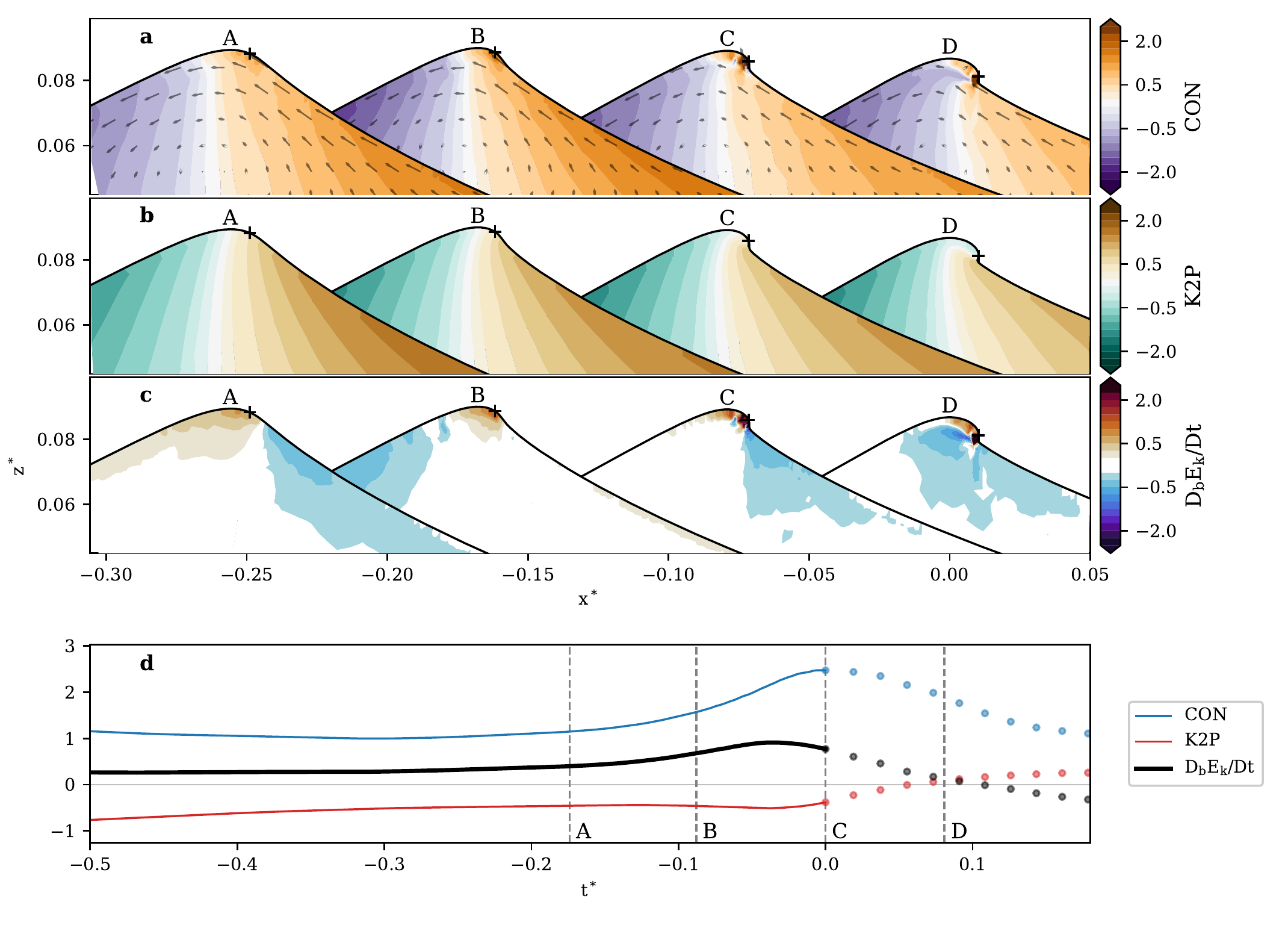}
  \caption{As for figure \ref{fig:balance-slice-NB1} showing the B1 crest. Periods of wave breaking are indicated by the solid dots.}
\label{fig:balance-slice-B1}
\end{figure}

\begin{figure}
	\includegraphics[width=\textwidth]{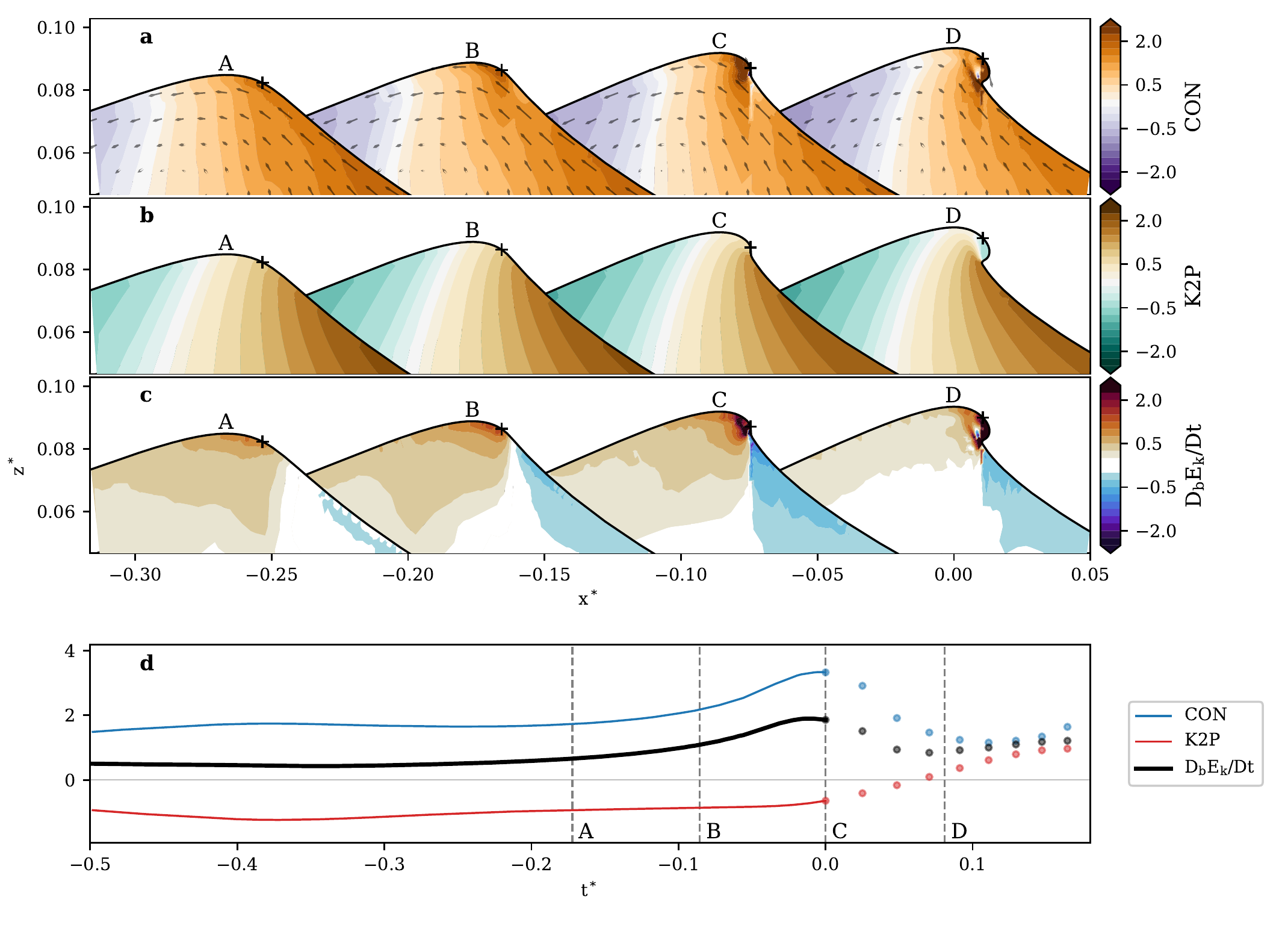}
  \caption{As for figure \ref{fig:balance-slice-NB1} showing the B2 crest. Periods of wave breaking are indicated by the solid dots.}
\label{fig:balance-slice-B2}
\end{figure}

The interplay between \con{} and \KP{} is examined in detail for the near-breaking crest NB1 (figure \ref{fig:balance-slice-NB1}) and breaking crests B1, B2 (figures \ref{fig:balance-slice-B1}-\ref{fig:balance-slice-B2}) by exploring their spatial variation within the evolving crests. The flux of kinetic energy within the crest is driven by the $\mathbf{u-b}$ vectors (panel b), which follow the approximate shape of the wave, decelerating as it moves upward and rearward from the forward side of the wave to the crest tip and then accelerating down the rear face. This flux leads to a convergence of kinetic energy on the forward side of the crest and divergence on the rear side. In the NB1 case, the local magnitude of the \con{} field is mostly offset by the \KP{} term (figure \ref{fig:balance-slice-NB1}c), so that the local rate of change of kinetic energy $D_b E_k / Dt$ is near-zero across most of the crest (figure \ref{fig:balance-slice-NB1}). 

In contrast, the \con{} fields in the B1 (figure \ref{fig:balance-slice-B1}) and B2 (figure \ref{fig:balance-slice-B2}) cases develop a local hotspot on the forward face of the crest tip, which continues to intensify up to breaking onset ($t^* = 0$). However, this is not offset by an equivalent hotspot in the local \KP{} field, whose characteristics are unchanged from the NB1 case. The imbalance between these two leading terms results in a corresponding intensification in the local rate of change $D_b E_k / Dt$.

Because of the large gradients in the evolving $D_b E_k / Dt$ field (figures \ref{fig:balance-slice-NB1}-\ref{fig:balance-slice-B2}) the temporal evolution of this and its contributing terms will be sensitive to the choice of the location \hotspot{} that is tracked within the crest. As described in section \ref{sec:ensemble}, we track the location of maximum $E_k$; this choice was made both for practical reasons as it can be applied in a laboratory setting and because the location varies smoothly in time. But it can also be seen to correspond closely with the location of maximum \con{} and $D_b E_k / Dt$ (figures \ref{fig:balance-slice-NB1}-\ref{fig:balance-slice-B2}) and therefore captures the maximum intensity of the signal of interest. 

We quantify the degree to which $D_b E_k / Dt$ at \hotspot{} is representative of the energetics across the crest region by comparing these values with those integrated over a larger region of interest (appendix \ref{sec:sensitivity}) and find that the rapid increase in $D_b E_k / Dt$ and \con{} remains observable if the values are integrated over the top $20 \%$ of the crest for the weakly breaking B1 case and over the top $50 \%$ for the stronger B2 case. However, if the energetic values are integrated over the full water depth this energetic signature is no longer evident. As well as validating our use of the local energetic quantities, these results also demonstrate the subtlety of this process that is not easily observed in bulk energy values. 

\subsection{Energy balance and breaking inception}

The key feature of the kinetic energy evolution leading up to breaking onset has been shown above to be a breakdown in the approximate equilibrium between the source of kinetic energy \con{} and the sink \KP{}, with friction becoming significant only near $t^*=0$.  This phenomenon is most clearly observed in the local values at the $E_k$ maxima, but is also evident to a lesser extent when integrating over sub-regions of the crest tip of various sizes (appendix \ref{sec:sensitivity}). From this we can conclude that the evolution of the local energetic quantities at \hotspot{} accurately characterises the broader dynamics of the crest tip region. 

\begin{figure}
    \includegraphics[width=\textwidth]{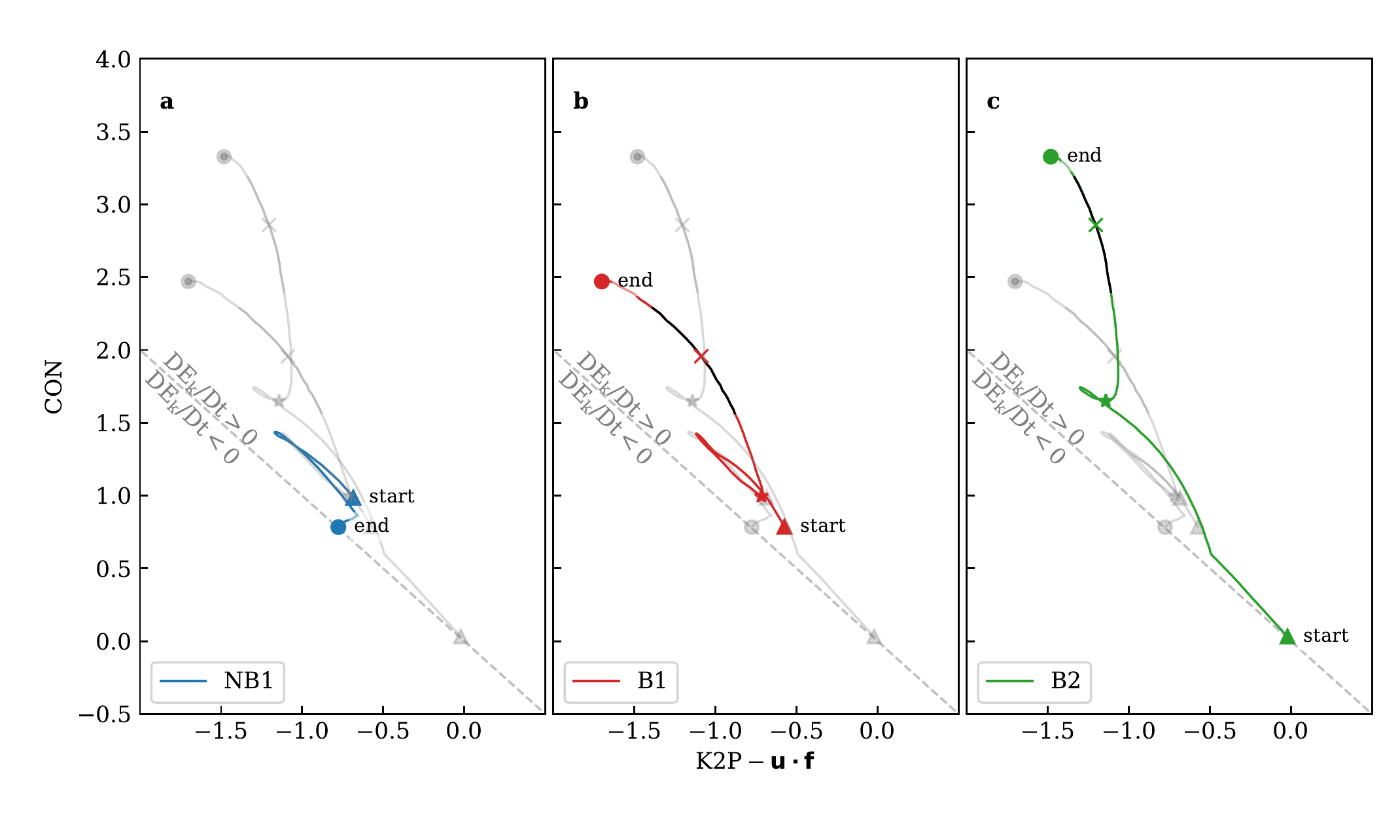}
 \caption{Relationship between the two leading terms in the $E_k$ balance equation (\ref{eq:DbEkDt-balance}): \KP{} $- \mathbf{u \cdot f}$ and \con{}, for the period $t^*=-1$ to $t^*=0$. Values above (below) the dashed line indicate an imbalance between these terms which leads to an increase (decrease) in $D_b E_k / Dt$. The energetic inception time ($\star$) and the kinematic inception time (i.e. when $B=B_{th}$, indicate by $\times$) for breaking crests are annotated, with the superposed black lines indicating the period for which $0.83 < B < 0.88$. An animated version of this figure is provided as supplementary material. }
 \label{fig:div-vs-rho}
\end{figure}

The relationship between these terms during the final wave period leading up to $t^*=0$ is further examined in figure \ref{fig:div-vs-rho}, an animated version of which is also provided as supplementary material. Here, the grey dashed line denotes an equal balance between the source (\con{}) and sink (\KP{}, $\mathbf{u \cdot f}$) terms with any departure above (below) this line indicating a resultant increase (decrease) in $E_k$ (ignoring the insignificant contribution of  the surface tension term in (\ref{eq:DbEkDt-balance})).

In the NB1 near-breaking crest, the initial convergence of kinetic energy is mostly offset by the sink terms, with the magnitude of these terms eventually peaking as the trajectory of the line reverses direction. Throughout this process, the magnitude of each of these terms increases at a similar rate, which keeps the distance from the $D_b E_k / Dt = 0$ line consistent, so that the growth rate of kinetic energy stays within reasonable bounds. 

The breaking case B1 initially follows a near-identical trajectory, with the convergence of kinetic energy sufficiently offset by the sink terms to ensure a steady increase in $E_k$. However, at some time after the magnitude of \con{} begins to decrease, it suddenly experiences a rapid increase (indicated by the $\star$ symbol) that is not balanced by a corresponding increase in the magnitude of the sink terms. As a consequence $D_b E_k / Dt$ also rapidly grows up to breaking onset. A similar evolution is also observed for the stronger B2 breaking case. Here, the imbalance between source and sink terms is larger and so the trajectory is displaced further from the $D_b E_k / Dt = 0$ line, indicating a faster growth in kinetic energy. As in the B1 case, the magnitude of \con{} begins to decrease before a striking deflection is seen and \con{} grows rapidly, driving a subsequent rapid growth in $D_b E_k / Dt$. 

\begin{figure}
    \begin{subfigure}[b]{0.5\textwidth}
         \centering
         \includegraphics[height=0.375\textheight]{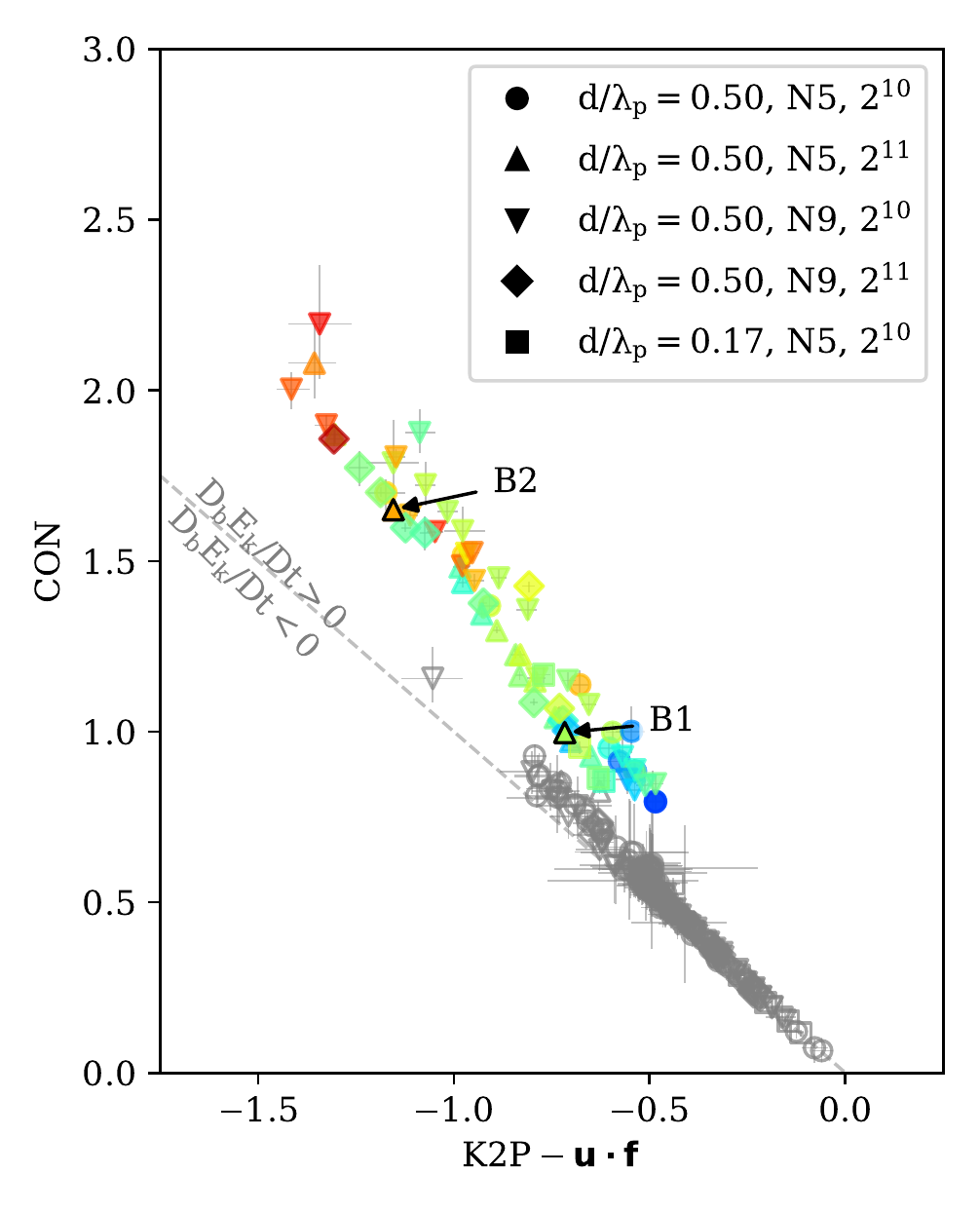}
    \end{subfigure}
    \begin{subfigure}[b]{0.5\textwidth}
         \centering
         \includegraphics[height=0.37\textheight]{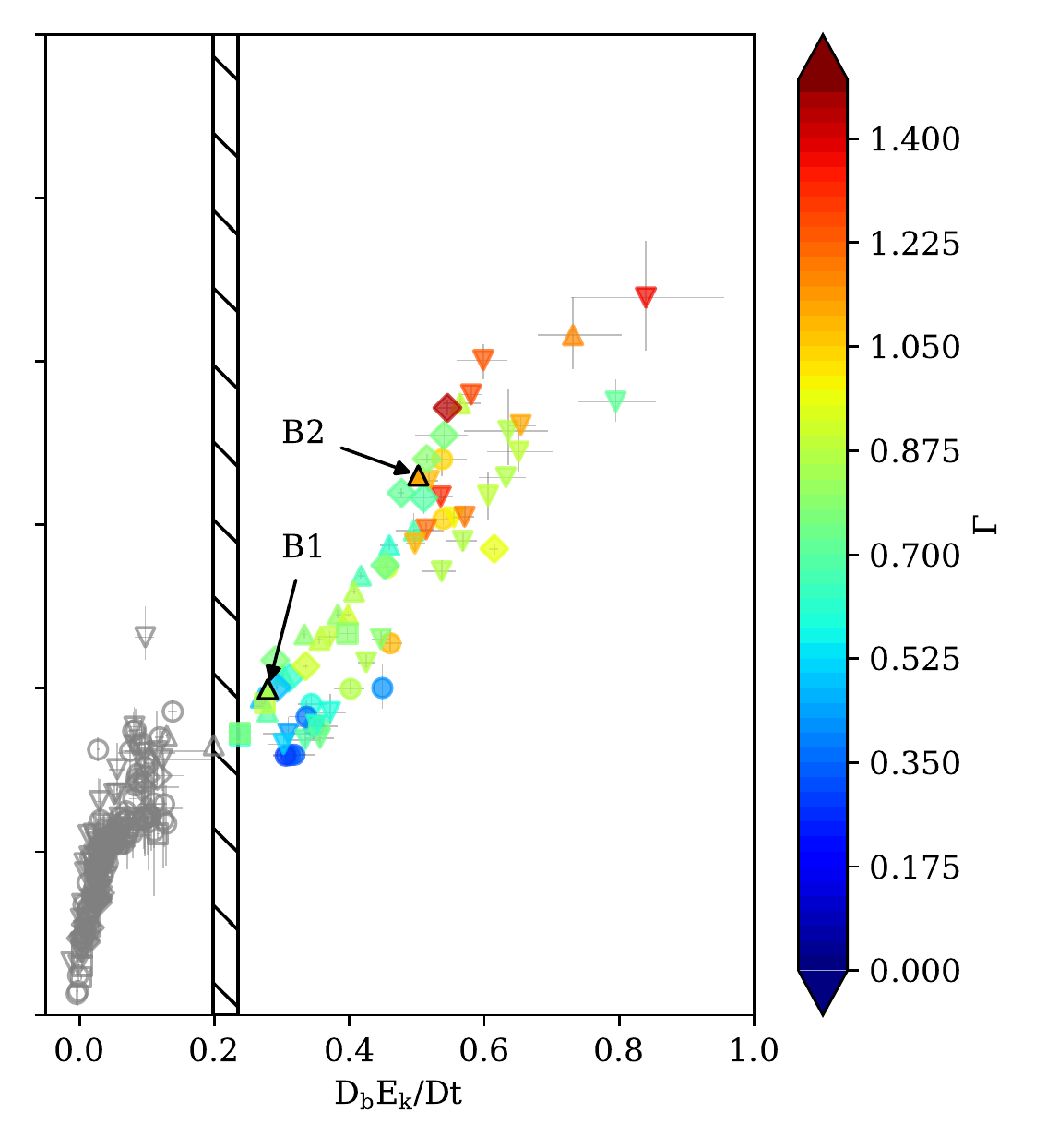}
    \end{subfigure}
  \caption{(left) The magnitude of \con as a function of (left) $K2P - \mathbf{u \cdot f}$ and (right) $D_b E_k / Dt$, at the time that the critical point $p^{\star}$ occurs. The $5 \%$ and $95 \%$ confidence intervals are also shown. Breaking crests are coloured by the breaking strength indicator $\Gamma$, and non-breaking crests are coloured grey. The $D_b E_k / Dt$ threshold separating non-breaking and breaking crests is shown by the hatched region.}
\label{fig:div-vs-rho-summary}
\end{figure}

In order to test the existence of this inflection point for all cases in our ensemble, we define the critical point $p^\star$ as the final local minimum of the parametric curve $(x,y) = ([K2P - \mathbf{u \cdot f}], \; CON)$ that occurs before $t^*=0$. Without exception, we find that all breaking crests feature a rapid increase in \con{} commencing at the critical point $p^\star$ (figure \ref{fig:div-vs-rho-summary}, solid symbols), which subsequently leads to a rapid increase in $D_b E_k / Dt$ up to breaking onset. This generic feature of the crest evolution therefore represents a critical energy imbalance between the kinetic energy source and sink terms.

The occurrence of this critical point is not in itself a sufficient condition for breaking. We observe that small inflections do occur for some non-breaking waves, although these are not followed by a rapid increase in \con{}. In these non-breaking cases (figure \ref{fig:div-vs-rho-summary}, grey symbols), the magnitude of the \con{} term at the critical point may even exceed that of some of the breaking crests and by itself this is clearly not a distinguishing feature between non-breaking and breaking waves. However, the magnitude of $D_b E_k / Dt$ at this critical point does distinguish between the two classes (figure \ref{fig:div-vs-rho-summary} right). For non-breaking crests, $D_b E_k / Dt$ is mostly near-zero whereas $D_b E_k / Dt$ increases near-linearly with increasing \con{} for breaking crests. Moreover, the values of $D_b E_k / Dt$ for non-breaking and breaking crests are clearly separated by a threshold region $D_b E_k / Dt = [0.198, 0.235]$. Therefore, we see that the distinguishing energetic feature separating breaking and non-breaking crests is the occurrence of a critical point in the counterbalance between $E_k$ source and sink terms when $ D_b E_k / Dt \geq 0.235$. Beyond this threshold value of the kinetic energy growth rate, the sink terms cannot absorb the continuing increase in $E_k$ and the wave passes an energetic point of no return that culminates in breaking onset. This process represents an energetic indicator of breaking inception and we hereafter label this critical point in the crest evolution the energetic signature for breaking inception. 

\subsection{Features of the energetic signature for breaking inception}

We now explore the prognostic characteristics of this new energetic signature for breaking inception and look first at the breaking strength. Up to this point we have characterised this through a visual examination of the interface evolution of our representative breaking crests B1 and B2 (e.g. figure \ref{fig:KE-slice}), with the B2 crest exhibiting more extensive overturning. To quantify this assessment, we use the breaking strength parameter \citep{DerakhtiMorteza2018Ptbs}
\begin{equation}\label{eq: GammaB_def}
    \Gamma = T \left. \frac{D_b B}{Dt}\right|_{B_\mathrm{th}}.
\end{equation}
While other methods of defining the breaking strength exist \citep[e.g.][]{drazen2008}, we utilise (\ref{eq: GammaB_def}) as it is conveniently formulated with the same local energetic quantities that we are investigating. The parameter $D_b B / Dt$ is calculated as the average rate of change over the time that $0.83 < B < 0.88$ and the wave period $T$ is defined from the deep water relationship using the crest zero-crossing wavelength $\lambda_c$. The value of $\Gamma$ for the B1 and B2 crests is $0.85$ and $1.1$ respectively, which aligns with our initial qualitative assessment of breaking strength. In figure \ref{fig:div-vs-rho-summary}, where all breaking crests are coloured by the magnitude of $\Gamma$, we see that the largest $\Gamma$ values are associated with the largest values of \con{}, the largest imbalance between \con{} and $\rho g w - \mathbf{u \cdot f}$ and the largest $D_b E_k / Dt$ values at the time that the energetic inception signature occurs. These results demonstrate that the magnitude of these terms at the instant of the energetic inception signature give an indication of the strength of the subsequent breaking event. 

\begin{figure}
     \includegraphics[width=\textwidth]{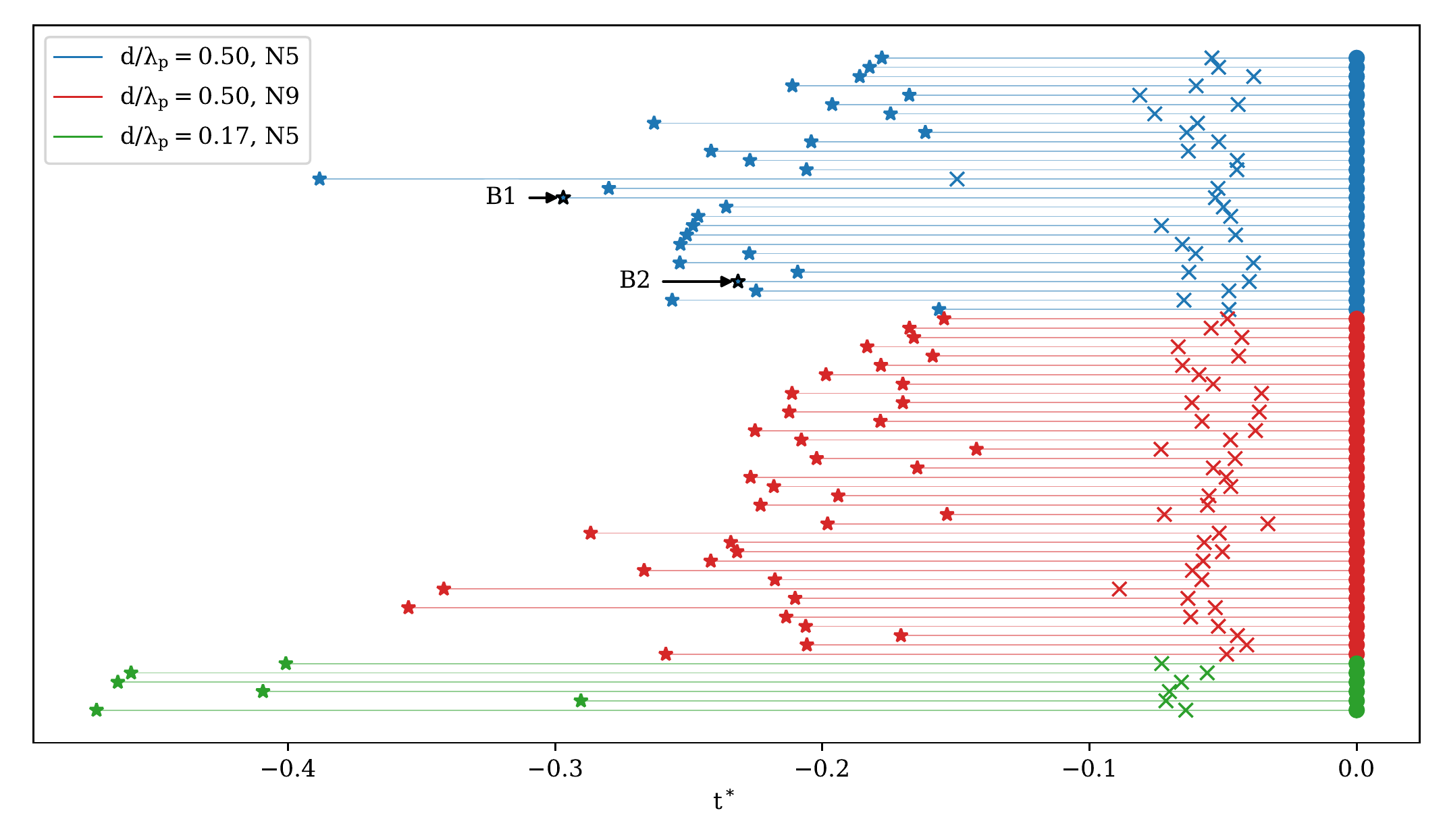}
 \caption{The energetic inception time ($\star$) and the kinematic inception time ($\times$) for all breaking crests in our ensemble, relative to the time of breaking onset $t^*=0$. Crests are grouped by wave packet size and water depth and the representative breaking crests B1 and B2 discussed in the text are annotated.}
 \label{fig:inception-time}
\end{figure}

The timing of the energetic inception signature is also of particular interest as the identification of breaking inception provides advance warning of breaking onset. The instant at which $B=B_{th}$, which we refer to as the kinematic inception threshold, is shown for the B1 and B2 cases in figure \ref{fig:div-vs-rho} by the $\times$ symbol. The period over which $0.83 < B < 0.88$ has also been coloured black to provide an indication of the rate of change in $B$ around this time. The timing of the energetic inception signature ($\star$) is clearly separated from the kinematic inception threshold and occurs earlier in both examples. This is seen to be the case for all breaking crests in our ensemble (figure \ref{fig:inception-time}). While the kinematic inception threshold consistently occurs around $0.05 - 0.1$ wave periods prior to breaking onset regardless of wave packet size or water depth, the energetic inception signature occurs much earlier, up to $0.4$ wave periods prior to breaking onset for our deep water crests and up to $0.7$ wave periods prior for our shallow water cases. 

\section{Discussion and conclusions}

Using an ensemble of breaking, near-breaking and non-breaking wave crests simulated with a numerical wave tank, we have examined the evolution of the kinetic energy balance as crests transition from growth to decay. Our results provide new details on the energetic processes leading up to the onset of breaking and the key difference between non-breaking and breaking crest evolution. 

Relative to the crest motion, the kinetic energy field is driven by a flux velocity that moves upward and rearward to the crest tip before descending down the rearward face of the wave. On the forward side of the crest, this flux drives a net convergence of kinetic energy, the majority of which is converted to potential energy as the fluid is lifted. In non-breaking crests, the net rate of change of kinetic energy only modestly varies between growth and decay as this convergence of kinetic energy is nearly offset by conversion to potential energy. In breaking crests the rate of kinetic energy convergence is significantly larger than in non-breaking crests, but the rate of conversion to potential energy is similar for all crests. This imbalance leads to a net increase in total kinetic energy, which continues up to breaking onset. Of the remaining terms in the kinetic energy balance equation (\ref{eq:DbEkDt-balance}), surface tension plays a negligible role and viscous dissipation is only significant on a local scale at the crest tip. 

These energetic processes are highly localised and, while the general characteristics can be seen in bulk energetic values, the full detail is only revealed when observing the local values. The maximum kinetic energy, as well as the largest values of kinetic energy convergence and rate of change, were observed to occur on the forward face of the crest tip. We confirmed that the evolution of these terms at this location is representative of the wider crest region by comparing these with the equivalent values integrated over various sub-regions of the crest. We found that the local variability at this hotspot is still evident even when integrating over the top $20 \%$ of the crest for weaker breaking cases and the top $50 \%$ for stronger breaking cases. 

These local values highlight the energetic signature that distinguishes breaking crests from their non-breaking counterparts. Throughout the evolution of a non-breaking crest, the rate of change of kinetic energy at the crest tip is bounded by the interplay between the source and sink terms, with the threshold range separating non-breaking and breaking crests in our ensemble found to be $D_b E_k / Dt = [0.198,0.235]$. But in a breaking crest, this threshold is exceeded and any further increase in kinetic energy through convergence can no longer be offset by conversion to potential energy or dissipation through friction. The result is an irreversible and rapid increase in kinetic energy that leads to breaking onset. 

Our results show that this energetic signature is a robust indicator of breaking inception. For our ensemble, this typically occurs around $0.25$ wave periods prior to breaking onset, but up to $0.7$ wave periods prior for the shallow water cases that we investigated with $d/\lambda_p = 0.2$. Of fundamental interest is that this energetic inception signature occurs significantly earlier than the kinematic inception threshold based on the transition of $B$ through the value $B_{th} = 0.855$. 

A number of questions are left for future studies. We anticipate that the energetic signature of breaking inception presented here will be a consistent feature regardless of wave packet type, water depth or wind forcing, but our ensemble has so far explored only a subset of these variables and further investigation is needed before this can be confirmed. A full energetic explanation for the existence of the kinematic inception threshold $B_{th}$ also remains unresolved, particularly as the time of this threshold is clearly distinct from our energetic inception signature presented here. Finally, while this study has focused on the time period leading up to breaking onset, our results also indicate a clear correlation between the breaking inception point and the strength of the breaking event, which has implications for the amount of energy dissipated. While we show a strong relationship between the kinetic energy convergence, the rate of change of kinetic energy and the breaking strength parameter $\Gamma$, we leave a full examination of this result for future work. 

\backsection[Acknowledgements]{This research was supported by the Australian Government's National Collaborative Research Infrastructure Strategy (NCRIS), with access to computational resources provided by the National Computing Infrastructure through the National Computational Merit Allocation Scheme and the University of New South Wales (UNSW) Resource Allocation Scheme. Additional computational resources were provided by the computational cluster Katana supported by Research Technology Services at UNSW Sydney.}

\backsection[Funding]{D.B. is supported by an Australian Government Research Training Program (RTP) Scholarship.} 

\backsection[Declaration of interests]{The authors report no conflict of interest.}

\backsection[Author ORCIDs]{Daniel G. Boettger https://orcid.org/0000-0002-9180-1481; Shane R.  Keating https://orcid.org/0000-0002-6817-925X; Michael L. Banner https://orcid.org/0000-0002-0799-5341; Russel P. Morison https://orcid.org/0000-0003-0721-1559; Xavier Barth\'el\'emy https://orcid.org/0000-0003-0285-0116.}

\backsection[Author contributions]{D.B. performed all aspects of the computations with technical support from X.B. M.B. coordinated the scientific effort in close collaboration with D.B., S.K. and R.M. D.B. drafted this paper, with significant technical and intellectual input on the analysis and interpretation of the results from S.K., M.B., R.M. and X.B.}

\bibliographystyle{jfm}
\bibliography{reference-library}

\appendix
\section{Convergence of the simulations with increasing resolution}\label{sec:Verification}

The numerical wave tank is configured to efficiently focus high grid refinement only where it is required to resolve the wave energetics. We confirm that the total simulation energy converges as a function of maximum grid refinement $2^n$  by performing a series of simulations with varying refinement levels but identical tank and paddle forcing settings. The total energy is calculated within a control volume covering the water phase but excluding the numerical sponge layer (figure \ref{fig:tank}). The total kinetic energy $\mathcal{K}$, potential energy $\mathcal{P}$ and their sum $\mathcal{E}$ are

\begin{equation}\label{eq:verification-integral}
    \mathcal{E}(t) = \mathcal{K}(t) + \mathcal{P}(t) = \int_0^{x_s} \int_{-H}^{\eta(x,t)} E \; dx \; dz \; + \int_0^t \left( \left. \int_{-H}^\eta uE \right|_0 \; dz \; - \left. \int_{-H}^\eta uE \right|_{x_s} \; dz \right) \;  dt
\end{equation}
where the integral limits extend from the bottom of the numerical wave tank $z=-H$ to the interface $\eta(x,t)$ and horizontally from the paddle boundary $x=0$ to the commencement of the sponge layer $x=x_s$. The flux of any energy in or out of the control volume is captured by the final two terms. 

\begin{figure}
  \centerline{\includegraphics[width=\textwidth]{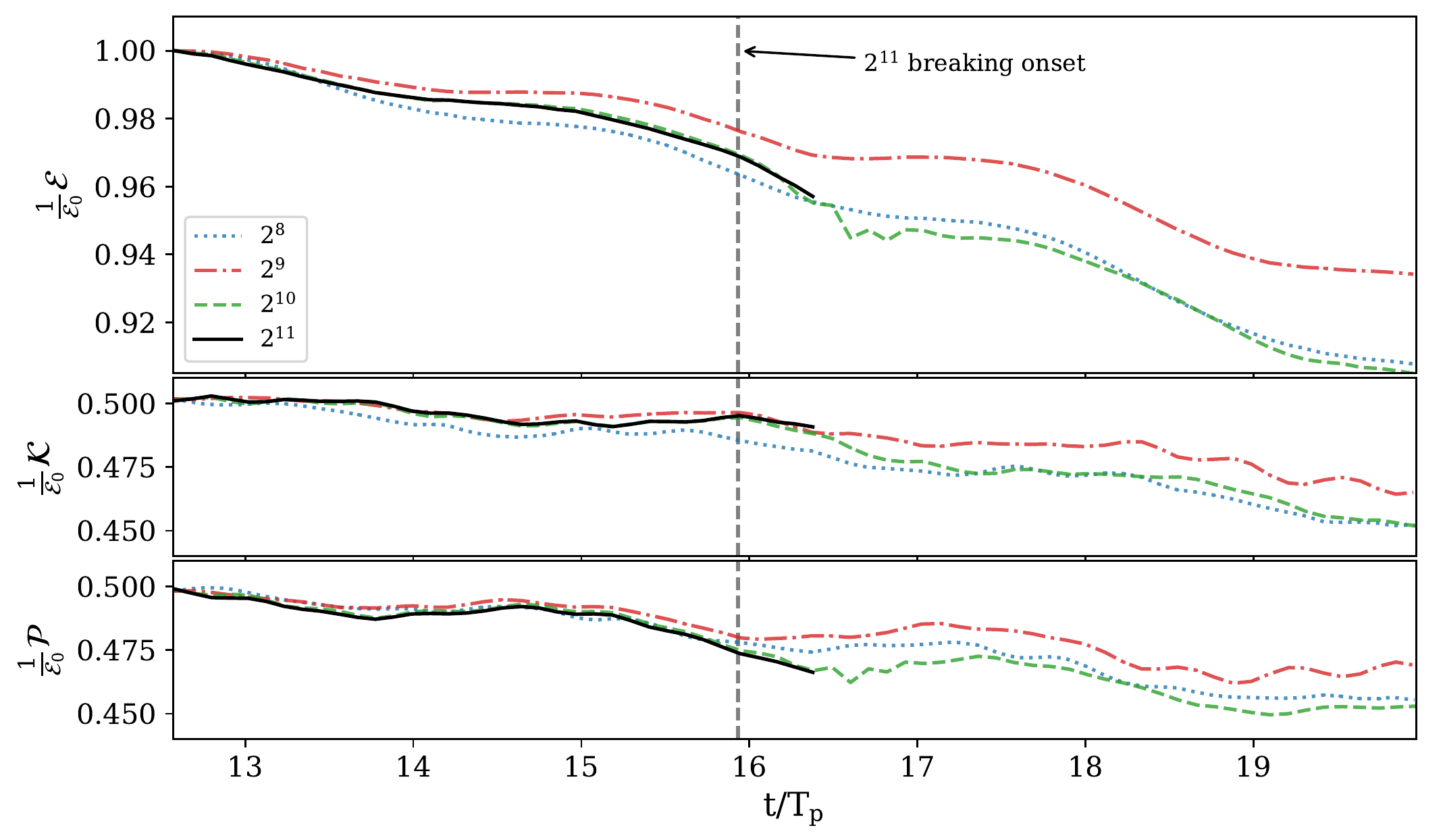}}
  \caption{The integrated energy of the water phase (excluding the sponge layer) for the simulation from which the near-breaking B2 crest has been taken (black) and for identical simulations with lower maximum grid refinement levels $2^n$. The total $\mathcal{E}$ (top), kinetic $\mathcal{K}$ (middle) and potential energy $\mathcal{P}$ (bottom) are shown. All values are normalised by $\mathcal{E}_0$: the total energy at the time which the wave packet has fully entered the numerical wave tank. Breaking onset time for the $2^{11}$ B2 crest is indicated by the dashed grey line.}
\label{fig:verification}
\end{figure}

In figure \ref{fig:verification} the total $\mathcal{E}$, $\mathcal{K}$ and $\mathcal{P}$ is shown for the simulation from which the breaking B2 crest has been taken, as well as simulations with identical paddle amplitude settings but smaller grid refinement. Energy values are normalised by the total energy $\mathcal{E}$ at simulation time $t/T_p = 12.5$, when the wave packet has fully entered the simulation domain and flux into the control volume is near-zero. This simulation is forced by one of the larger paddle amplitudes in our ensemble and we see similar results for other simulations.

For the period leading up to breaking onset for the B2 crest, the energy levels at each refinement level are similar. The equipartitioning of $\mathcal{K}$ and $\mathcal{P}$ within the wave packet is evident, with the  oscillations in these terms indicative of the conversion of energy within the wave packet. 

The onset of breaking is followed by the decrease in $\mathcal{E}$ and $\mathcal{K}$, which can be seen for both the $2^{10}$ and $2^{11}$ cases. Breaking is not observed for the lower refinement levels and the decrease in $\mathcal{E}$ is due to viscous and numerical dissipation only. The $\mathcal{E}$, $\mathcal{K}$ and $\mathcal{P}$ results converge for grid refinement levels $2^{10}$ and $2^{11}$ both before and after breaking onset, indicating that the energetic processes leading to breaking are sufficiently resolved at these resolutions. Our confidence that the local energetic processes are sufficiently resolved is further reinforced by the convergence of the $2^{10}$ and $2^{11}$ results presented in section \ref{sec:energetics}.

\section{Sensitivity of results to sampling location}\label{sec:sensitivity}

\begin{figure}
    \centering
    \includegraphics[width=\textwidth]{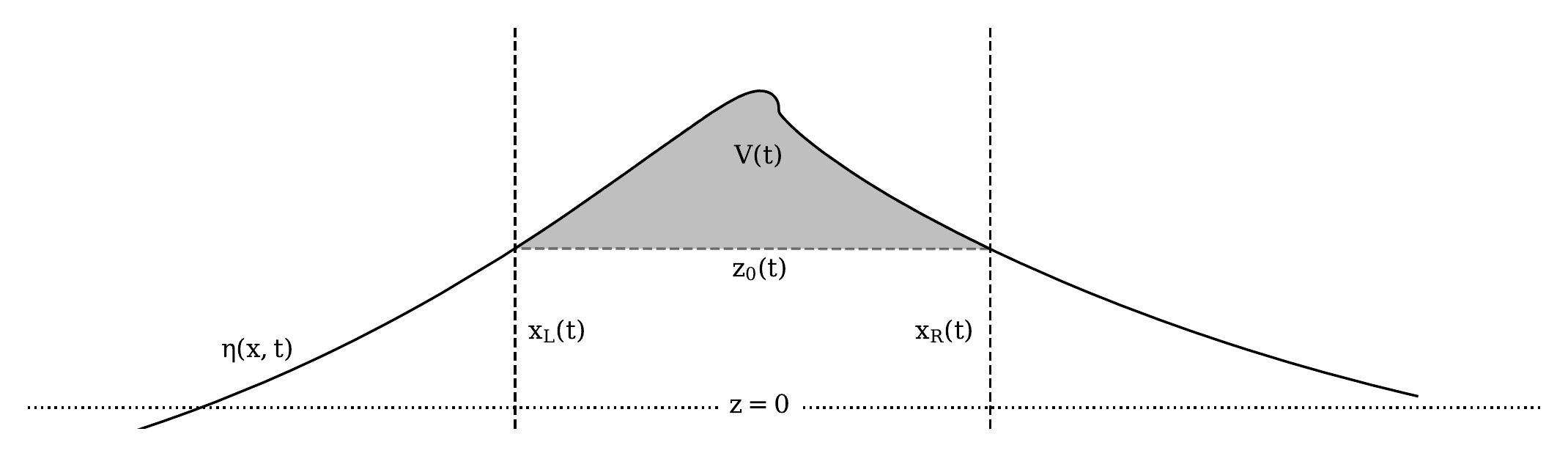}
    \caption{Schematic of the control volume $V(t)$ bounded at the top by the interface $\eta(x,t)$, at the bottom by the horizontal plane $z_0(t)$ and on the left and right by the vertical planes $x_L(t)$ and $x_R(t)$. The free variables $z_0$, $x_L$ and $x_R$ are set to examine a chosen sub-region of the crest.}
    \label{fig:volume-schematic}
\end{figure}

The rate of change of a scalar quantity $f$ integrated over a moving and deforming control volume $V$ with bounding surface $S$ is calculated using the Reynolds Transport Theorem,
\begin{equation}\label{eq:RTT}
   \frac{D_v \mathcal{F}} {Dt} = \frac{d}{d t} \int_{V(t)}  f \; dV \; = \int_{V(t)} \frac{\partial  f}{\partial t} \; dV + \int_{S(t)} \mathbf{b} \cdot \mathbf{n} f \; dS,
\end{equation}
where $\mathbf{b}$ is the local velocity of $S$. We define a control volume that moves and deforms to follow the crest evolution, bounded at the top and sides by the interface $\eta(\mathbf{x}, t)$ and at the bottom by a horizontal slice through $z=z_0(t)$ that intersects the interface at the locations $x_L(t)$, $x_R(t)$ (figure \ref{fig:volume-schematic}). In this case (\ref{eq:RTT}) can be formulated as
\begin{equation}\label{eq:RTT-crest}
    \frac{D_v \mathcal{F}} {Dt} = 
    \int _{x_L}^{x_R}\int_{z_0}^\eta \frac{\partial f}{\partial t} \; dx \; dz \; +
     \left. \int_{x_L}^{x_R} \mathbf{b} \cdot \mathbf{n} f \right|_{z_0} \; dS \; +
     \left. \int_{x_L}^{x_R} \mathbf{b} \cdot \mathbf{n} f \right|_{\eta} \; dS \;,
\end{equation}
where the first surface integral encompasses the part of the control surface within the crest and the second is the remaining control surface along the interface. 

The integrated rate of change of $E_k$ is found by substituting (\ref{eq:Ek-balance}) for $f$ in (\ref{eq:RTT-crest})
\begin{multline}
   \frac{D_v \mathcal{K}} {Dt} =
    \int _{x_L}^{x_R}\int_{z_0}^\eta \left( -\nabla \cdot \mathbf{u} \left( p + E_k \right) - \rho g w + \mathbf{u \cdot f} + \mathbf{u \cdot n} \sigma \kappa \delta_s \right) \; dx \; dz \; + \\
     \left. \int_{x_L}^{x_R}  \mathbf{b} \cdot \mathbf{n} E_k \right|_{z_0} \; dS \; +
     \left. \int_{x_L}^{x_R} \mathbf{b} \cdot \mathbf{n} E_k \right|_{\eta} \; dS \;.
\end{multline}
With application of the divergence theorem this becomes
\begin{multline}\label{eq:KE-RTT-surface}
   \frac{D_v \mathcal{K}} {Dt} =
    \int_{x_L}^{x_R} \int_{z_0}^\eta \left( - \rho g w + \mathbf{u \cdot f} + \mathbf{u \cdot n} \sigma \kappa \delta_s \right) \; dx \; dz \; +
    \left. \int_{x_L}^{x_R} \mathbf{u} \cdot \mathbf{n} p \right|_{z_0} \; dx \; +
    \left. \int_{x_L}^{x_R} \mathbf{u} \cdot \mathbf{n} p \right|_{\eta} \; dS \; + \\
     \left. \int_{x_L}^{x_R} (\mathbf{u - b}) \cdot \mathbf{n} E_k \right|_{z_0} \; dx \; +
     \left. \int_{x_L}^{x_R} (\mathbf{u - b}) \cdot \mathbf{n} E_k \right|_{\eta} \; dS \;,
\end{multline}
where at the interface $\mathbf{u \cdot n = b \cdot n}$ so that the last term cancels and the remaining surface integrals can be expressed as volume integrals to arrive at
\begin{equation}
    \frac{D_v \mathcal{K}}{D t} = \int_{x_L}^{x_R} \int_{z_0}^\eta \left( - \nabla \cdot \left(\mathbf{u} p + [\mathbf{u-b}]E_k \right) - \rho g w + \mathbf{u \cdot f} + \mathbf{u \cdot n} \sigma \kappa \delta_s \right) \; dx \; dz.
    \label{eq:KE-RTT}
\end{equation}

The balance equation (\ref{eq:KE-RTT}) has a number of favourable properties. Firstly, it can be seen that if (\ref{eq:KE-RTT}) is applied to an arbitrarily small control volume the local kinetic energy balance equation (\ref{eq:DbEkDt-balance}) is recovered. In addition, as the air-water interface at $\eta$ is a material surface, the divergence of kinetic energy within the control volume is equal only to the relative flux of kinetic energy through the $z_0$ plane
\begin{equation}
    \int_{x_L}^{x_R} \int_{z_0}^\eta \nabla \cdot \left([\mathbf{u-b}]E_k \right) \; dx \; dz = \left. \int_{x_L}^{x_R} [\mathbf{u - b}] \cdot \mathbf{n} E_k \right|_{z_0} \; dx \;,
    \label{eq:flux}
\end{equation}
with the RHS of (\ref{eq:flux}) providing a more numerically convenient method for accurately calculating the divergence field within the control volume. 

To account for the temporal change in the size of the control volume, (\ref{eq:KE-RTT}) can alternatively be formulated as a volume-averaged quantity; however, this introduces an additional dilation term in (\ref{eq:KE-RTT}) that complicates the interpretation of the energy budget. The change in $V(t)$ is small (less than $10 \%$ over the final $0.2T_0$ prior to breaking onset for the $z_0 = 0.9 a(t)$ case) in comparison to the changes in the other terms. As our focus is on the relative magnitude of these terms in individual cases, the time-varying nature of $V(t)$ does not impact the results presented in this section. 

By adjusting the parameter $z_0$ the terms in (\ref{eq:KE-RTT}) can be examined for sub-regions of the crest tip of various sizes. We set $z_0$ as a fraction of the crest amplitude $a(t)$. The sub-region defined by $z_0 = 0.9 a(t)$ is the smallest control volume that fully encompasses the crest bulge and the region of large $E_k$ values around the crest tip (figure \ref{fig:KE-slice}). 

\begin{figure}
    \centering
    \includegraphics[width=\textwidth]{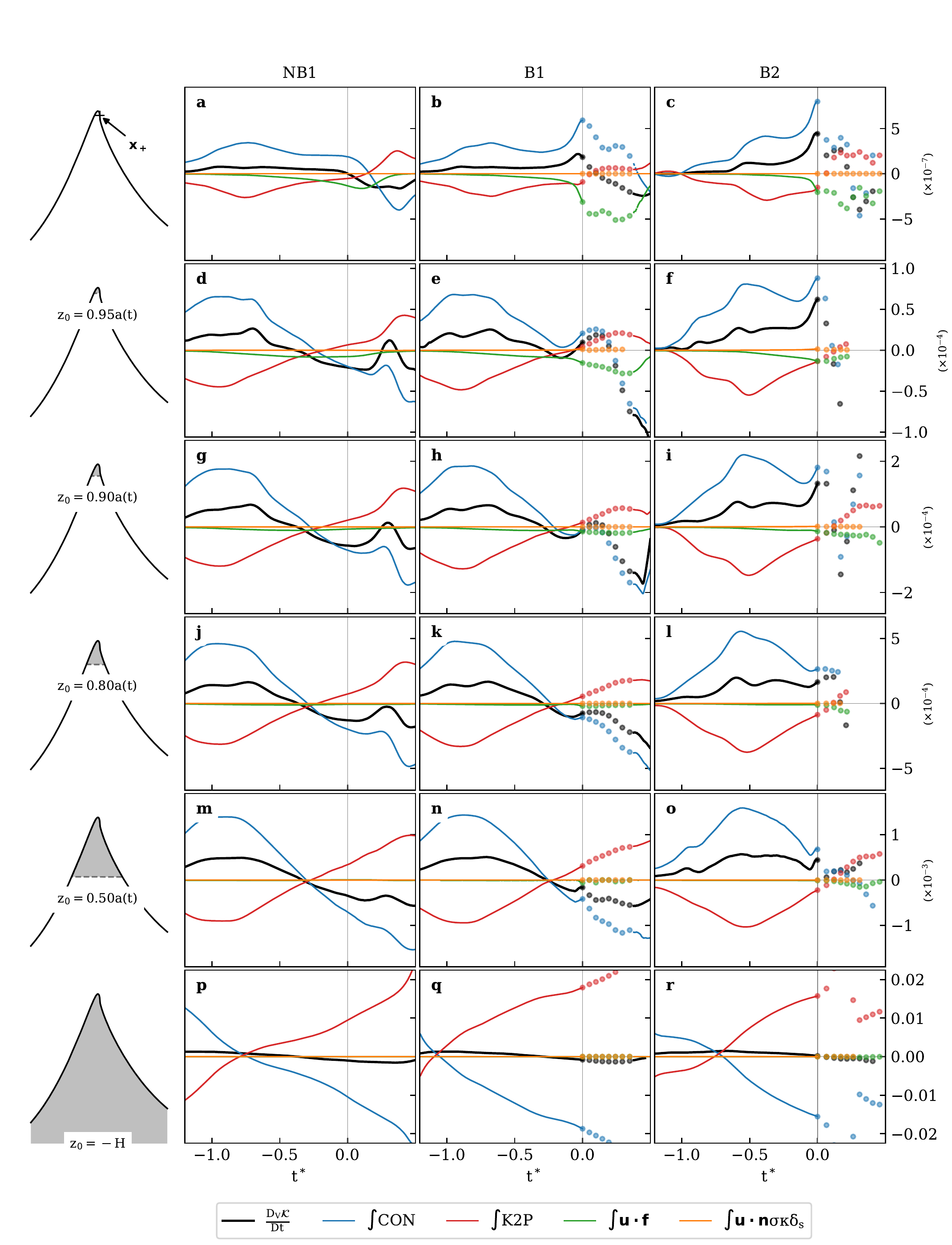}
    \caption{Evolution of the rate of change of the integrated kinetic energy $\mathcal{K}$ and its contributing terms from (\ref{eq:KE-RTT}) for the representative near-breaking (NB1) and breaking (B1, B2) crests. Positive (negative) values indicate a source (sink) of $\mathcal{K}$. Periods of wave breaking are indicated by the solid dots. The top row shows the local values taken at the location \hotspot{} (figure \ref{fig:local-energetics}) and multiplied by the model cell volume. Each remaining row displays values integrated over a control volume that encompasses increasing amounts of the crest as shown by each crest schematic (left).}
    \label{fig:integrated-energetics}
\end{figure}

In figure \ref{fig:integrated-energetics}, we compare the evolution of $D_b E_k / D t$ and its components from figure \ref{fig:local-energetics} to the integrated values obtained from (\ref{eq:KE-RTT}) for a range of control volumes that vary in size from the top $5 \%$ ($z_0 = 0.95a(t)$) to the top $50\%$ ($z_0 = 0.5a(t)$) of the evolving crest. The local $D_b E_k / D t$ values are converted to the same units as (\ref{eq:KE-RTT}) by multiplying by the grid cell volume, which is a constant value as \hotspot{} is located in the high-resolution interface region. For brevity we use a single integral symbol to refer to these integrated terms (e.g. \intcon{})

An initial observation from figure \ref{fig:integrated-energetics} is that the evolution of all terms is relatively consistent as the region of interest is increased in size from a point location (top row) to a large control volume (bottom). But in relation to the key findings from this study, the features of most interest are the rapid increase in $D_v \mathcal{K} / Dt$ (black line) and \intcon{} (blue) just prior to $t^*=0$ that is observed for the breaking crests B1 and B2. As the size of the control volume increases, the relative magnitude of this signal is diminished, but remains observable in the $z_0 = 0.8a(t)$ control volume for the B1 case (figure \ref{fig:integrated-energetics}n) and in the $z_0 = 0.5a(t)$ control volume for the stronger B2 case (figure \ref{fig:integrated-energetics}r). In contrast, for all three representative crests the \intKP{} and \intcon{} values are in close balance when integrated over the full water depth (figure \ref{fig:integrated-energetics}s,t,u) and the rate of change of $\mathcal{K}$ is consequently near-zero. 


\end{document}